\documentclass[longauth]{aa}  
\usepackage{graphicx}
\usepackage{txfonts}
\usepackage{multicol}
\usepackage{float}
\usepackage{subcaption}
\usepackage{multirow}
\usepackage{xcolor}

\usepackage{siunitx}
\usepackage{svg}
\usepackage{hyperref}

\begin{document} 

\authorrunning{N. Pourré et al.}

%   \title{Direct confirmation of a Gaia brown dwarf}
%\title{Bringing Gaia brown dwarfs to light with VLTI/GRAVITY \thanks{Based on observations collected at the European Southern Observatory under ESO programmes 1104.C-0651 and 0110.C-0182.}}

\title{High contrast at short separation with VLTI/GRAVITY: Bringing Gaia companions to light \thanks{Based on observations collected at the European Southern Observatory under ESO programs 1104.C-0651, 0110.C-0182 (GTO NAOMI), 60.A-9102 (GRAVITY+ commissioning run) and 0112.C-2396(C)}}

\author{ 
 N.~Pourr\'e\inst{\ref{ipag}}\thanks{Corresponding author:\newline N. Pourré (email: \texttt{nicolas.pourre@univ-grenoble-alpes.fr})}
 \and T.~O.~Winterhalder\inst{\ref{esog}}
 \and J.-B.~Le~Bouquin\inst{\ref{ipag}}
 \and S.~Lacour\inst{\ref{lesia},\ref{esog}}
 \and A.~Bidot\inst{\ref{ipag}}
 \and M.~Nowak\inst{\ref{cam}}
 \and A.-L.~Maire\inst{\ref{ipag}}
 \and D.~Mouillet\inst{\ref{ipag}}
 \and C.~Babusiaux\inst{\ref{ipag}}
 \and J.~Woillez\inst{\ref{esog}}
 \and R.~Abuter\inst{\ref{esog}}
 \and A.~Amorim\inst{\ref{lisboa},\ref{centra}}
 \and R.~Asensio-Torres\inst{\ref{mpia}}
 \and W.~O.~Balmer\inst{\ref{jhupa},\ref{stsci}}
 \and M.~Benisty\inst{\ref{ipag}}
 \and J.-P.~Berger\inst{\ref{ipag}}
 \and H.~Beust\inst{\ref{ipag}}
 \and S.~Blunt\inst{\ref{northwestern}}
 \and A.~Boccaletti\inst{\ref{lesia}}
 \and M.~Bonnefoy\inst{\ref{ipag}}
 \and H.~Bonnet\inst{\ref{esog}}
 \and M.~S.~Bordoni\inst{\ref{mpe}}
 \and G.~Bourdarot\inst{\ref{mpe}}
 \and W.~Brandner\inst{\ref{mpia}}
 \and F.~Cantalloube\inst{\ref{lam}}
 \and P.~Caselli \inst{\ref{mpe}}
 \and B.~Charnay\inst{\ref{lesia}}
 \and G.~Chauvin\inst{\ref{cotedazur}}
 \and A.~Chavez\inst{\ref{northwestern}}
 \and E.~Choquet\inst{\ref{lam}}
 \and V.~Christiaens\inst{\ref{liege}}
 \and Y.~Cl\'enet\inst{\ref{lesia}}
 \and V.~Coud\'e~du~Foresto\inst{\ref{lesia}}
 \and A.~Cridland\inst{\ref{leiden}}
 \and R.~Davies\inst{\ref{mpe}}
 \and D.~Defrère\inst{\ref{KULeuven}}
 \and R.~Dembet\inst{\ref{lesia}}
 \and J.~Dexter\inst{\ref{boulder}}
 \and A.~Drescher\inst{\ref{mpe}}
 \and G.~Duvert\inst{\ref{ipag}}
 \and A.~Eckart\inst{\ref{cologne},\ref{bonn}}
 \and F.~Eisenhauer\inst{\ref{mpe}}
 \and N.~M.~F\"orster Schreiber\inst{\ref{mpe}}
 \and P.~Garcia\inst{\ref{centra},\ref{porto}}
 \and R.~Garcia~Lopez\inst{\ref{dublin},\ref{mpia}}
 \and E.~Gendron\inst{\ref{lesia}}
 \and R.~Genzel\inst{\ref{mpe},\ref{ucb}}
 \and S.~Gillessen\inst{\ref{mpe}}
 \and J.~H.~Girard\inst{\ref{stsci}}
  \and F.~Gonte\inst{\ref{esog}}
 \and S.~Grant\inst{\ref{mpe}}
 \and X.~Haubois\inst{\ref{esoc}}
 \and G.~Hei\ss el\inst{\ref{actesa},\ref{lesia}}
 \and Th.~Henning\inst{\ref{mpia}}
 \and S.~Hinkley\inst{\ref{exeter}}
 \and S.~Hippler\inst{\ref{mpia}}
  \and S.~F.~H\"onig\inst{\ref{southamp}}
 \and M.~Houll\'e\inst{\ref{cotedazur}}
 \and Z.~Hubert\inst{\ref{ipag}}
 \and L.~Jocou\inst{\ref{ipag}}
 \and J.~Kammerer\inst{\ref{esog}}
 \and M.~Kenworthy\inst{\ref{leiden}}
 \and M.~Keppler\inst{\ref{mpia}}
 \and P.~Kervella\inst{\ref{lesia}}
 \and L.~Kreidberg\inst{\ref{mpia}}
 \and N.~T.~Kurtovic\inst{\ref{mpe}}
 \and A.-M.~Lagrange\inst{\ref{ipag},\ref{lesia}}
 \and V.~Lapeyr\`ere\inst{\ref{lesia}}
 \and D.~Lutz\inst{\ref{mpe}}
 \and F.~Mang\inst{\ref{mpe}}
 \and G.-D.~Marleau\inst{\ref{duisburg},\ref{tuebingen},\ref{bern},\ref{mpia}}
 \and A.~M\'erand\inst{\ref{esog}}
 \and F.~Millour\inst{\ref{cotedazur}}
 \and P.~Molli\`ere\inst{\ref{mpia}}
 \and J.~D.~Monnier\inst{\ref{umich}}
 \and C.~Mordasini\inst{\ref{bern}}
 \and E.~Nasedkin\inst{\ref{mpia}}
 \and S.~Oberti\inst{\ref{esog}}
 \and T.~Ott\inst{\ref{mpe}}
 \and G.~P.~P.~L.~Otten\inst{\ref{sinica}}
 \and C.~Paladini\inst{\ref{esoc}}
 \and T.~Paumard\inst{\ref{lesia}}
 \and K.~Perraut\inst{\ref{ipag}}
 \and G.~Perrin\inst{\ref{lesia}}
 \and O.~Pfuhl\inst{\ref{esog}}
 \and L.~Pueyo\inst{\ref{stsci}}
 \and D.~C.~Ribeiro\inst{\ref{mpe}}
 \and E.~Rickman\inst{\ref{esa}}
 \and Z.~Rustamkulov\inst{\ref{jhueps}}
 \and J.~Shangguan\inst{\ref{mpe}}
 \and T.~Shimizu \inst{\ref{mpe}}
 \and D.~Sing\inst{\ref{jhupa},\ref{jhueps}}
 \and F.~Soulez\inst{\ref{cral}}
 \and J.~Stadler\inst{\ref{mpa},\ref{origins}}
 \and T.~Stolker\inst{\ref{leiden}}
 \and O.~Straub\inst{\ref{origins}}
 \and C.~Straubmeier\inst{\ref{cologne}}
 \and E.~Sturm\inst{\ref{mpe}}
 \and C.~Sykes\inst{\ref{southamp}}
 \and L.~J.~Tacconi\inst{\ref{mpe}}
 \and E.~F.~van~Dishoeck\inst{\ref{leiden},\ref{mpe}}
 \and A.~Vigan\inst{\ref{lam}}
 \and F.~Vincent\inst{\ref{lesia}}
 \and S.~D.~von~Fellenberg\inst{\ref{bonn}}
 \and J.~J.~Wang\inst{\ref{northwestern}}
 \and F.~Widmann\inst{\ref{mpe}}
 \and S.~Yazici\inst{\ref{mpe}}
 \and  (GRAVITY Collaboration)
  \and J.~A.~Abad\inst{\ref{esog}}
 \and E.~Aller~Carpentier\inst{\ref{esog}}
 \and J.~Alonso\inst{\ref{esoc}}
 \and L.~Andolfato\inst{\ref{esog}}
 \and P.~Barriga\inst{\ref{esog}}
 \and J.-L.~Beuzit\inst{\ref{lam}}
 \and P.~Bourget\inst{\ref{esoc}}
 \and R.~Brast\inst{\ref{esog}}
 \and L.~Caniguante\inst{\ref{esoc}}
 \and E.~Cottalorda\inst{\ref{cotedazur}}
 \and P.~Darré\inst{\ref{esog}}
 \and B.~Delabre\inst{\ref{esog}}
 \and A.~Delboulbé\inst{\ref{ipag}}
 \and F.~Delplancke-Str\"obele\inst{\ref{esog}}
 \and R.~Donaldson\inst{\ref{esog}}
 \and R.~Dorn\inst{\ref{esog}}
 \and C.~Dupuy\inst{\ref{esog}}
 \and S.~Egner\inst{\ref{esog}}
 \and G.~Fischer\inst{\ref{esog}}
 \and C.~Frank\inst{\ref{esog}}
 \and E.~Fuenteseca\inst{\ref{esoc}}
 \and P.~Gitton\inst{\ref{esoc}}
 \and T.~Guerlet\inst{\ref{esog}}
 \and S.~Guieu\inst{\ref{ipag}}
 \and P.~Gutierrez\inst{\ref{esog}}
 \and P.~Haguenauer\inst{\ref{esog}}
 \and A.~Haimerl\inst{\ref{esog}}
 \and C.~T.~Heritier\inst{\ref{onera},\ref{lam}}
 \and S.~Huber\inst{\ref{esog}}
 \and N.~Hubin\inst{\ref{esog}}
 \and P.~Jolley\inst{\ref{esog}}
 \and J.-P.~Kirchbauer\inst{\ref{esog}}
 \and J.~Kolb\inst{\ref{esog}}
 \and J.~Kosmalski\inst{\ref{esog}}
 \and P.~Krempl\inst{\ref{KRP}}
 \and M.~Le~Louarn\inst{\ref{esog}}
 \and P.~Lilley\inst{\ref{esog}}
 \and B.~Lopez\inst{\ref{cotedazur}}
 \and Y.~Magnard\inst{\ref{ipag}}
 \and S.~Mclay\inst{\ref{esog}}
 \and A.~Meilland\inst{\ref{cotedazur}}
 \and A.~Meister\inst{\ref{esoc}}
 \and T.~Moulin\inst{\ref{ipag}}
 \and L.~Pasquini\inst{\ref{esog}}
 \and J.~Paufique\inst{\ref{esog}}
 \and I.~Percheron\inst{\ref{esog}}
 \and L.~Pettazzi\inst{\ref{esog}}
 \and D.~Phan\inst{\ref{esog}}
 \and W.~Pirani\inst{\ref{esog}}
 \and J.~Quentin\inst{\ref{esog}}
 \and A.~Rakich\inst{\ref{esog}}
 \and R.~Ridings\inst{\ref{esog}}
 \and J.~Reyes\inst{\ref{esog}}
 \and S.~Rochat\inst{\ref{ipag}}
 \and C.~Schmid\inst{\ref{esog}}
 \and N.~Schuhler\inst{\ref{esoc}}
 \and P.~Shchekaturov\inst{\ref{esog}}
 \and M.~Seidel\inst{\ref{esog}}
 \and C.~Soenke\inst{\ref{esog}}
 \and E.~Stadler\inst{\ref{ipag}}
 \and C.~Stephan\inst{\ref{esoc}}
 \and M.~Su\'arez\inst{\ref{esog}}
 \and M.~Todorovic\inst{\ref{esog}}
 \and G.~Valdes\inst{\ref{esoc}}
 \and C.~Verinaud\inst{\ref{esog}}
 \and G.~Zins\inst{\ref{esog}}
 \and S.~Z\'u\~niga-Fern\'andez\inst{\ref{esoc},\ref{univalp},\ref{nucleo}}, (NAOMI Collaboration)
 }
 
\institute{ 
   Univ. Grenoble Alpes, CNRS, IPAG, 38000 Grenoble, France
\label{ipag}      \and
   European Southern Observatory, Karl-Schwarzschild-Stra\ss e 2, 85748 Garching, Germany
\label{esog}      \and
   LESIA, Observatoire de Paris, PSL, CNRS, Sorbonne Universit\'e, Universit\'e de Paris, 5 place Janssen, 92195 Meudon, France
\label{lesia}      \and
   Institute of Astronomy, University of Cambridge, Madingley Road, Cambridge CB3 0HA, United Kingdom
\label{cam}      \and
   Universidade de Lisboa - Faculdade de Ci\^encias, Campo Grande, 1749-016 Lisboa, Portugal
\label{lisboa}      \and
   CENTRA - Centro de Astrof\' isica e Gravita\c c\~ao, IST, Universidade de Lisboa, 1049-001 Lisboa, Portugal
\label{centra}      \and
   Max Planck Institute for Astronomy, K\"onigstuhl 17, 69117 Heidelberg, Germany
\label{mpia}      \and
   Department of Physics \& Astronomy, Johns Hopkins University, 3400 N. Charles Street, Baltimore, MD 21218, USA
\label{jhupa}      \and
   Space Telescope Science Institute, 3700 San Martin Drive, Baltimore, MD 21218, USA
\label{stsci}      \and
   Center for Interdisciplinary Exploration and Research in Astrophysics (CIERA) and Department of Physics and Astronomy, Northwestern University, Evanston, IL 60208, USA
\label{northwestern}      \and
   Max Planck Institute for extraterrestrial Physics, Giessenbachstra\ss e~1, 85748 Garching, Germany
\label{mpe}      \and
   Aix Marseille Univ, CNRS, CNES, LAM, Marseille, France
\label{lam}      \and
Université Côte d'Azur, Observatoire de la Côte d'Azur, CNRS, Laboratoire Lagrange, Bd de l'Observatoire, CS 34229, 06304 Nice cedex 4, France
\label{cotedazur}      \and
  STAR Institute, Universit\'e de Li\`ege, All\'ee du Six Ao\^ut 19c, 4000 Li\`ege, Belgium
\label{liege}      \and
   Leiden Observatory, Leiden University, P.O. Box 9513, 2300 RA Leiden, The Netherlands
\label{leiden}      \and
   Department of Astrophysical \& Planetary Sciences, JILA, Duane Physics Bldg., 2000 Colorado Ave, University of Colorado, Boulder, CO 80309, USA
\label{boulder}      \and
 Institute of Astronomy, KU Leuven, Celestijnenlaan 200D, 3001 Leuven, Belgium
\label{KULeuven}    \and
   1.\ Institute of Physics, University of Cologne, Z\"ulpicher Stra\ss e 77, 50937 Cologne, Germany
\label{cologne}      \and
   Max Planck Institute for Radio Astronomy, Auf dem H\"ugel 69, 53121 Bonn, Germany
\label{bonn}      \and
   Universidade do Porto, Faculdade de Engenharia, Rua Dr.~RobertoRua Dr.~Roberto Frias, 4200-465 Porto, Portugal
\label{porto}      \and
   School of Physics, University College Dublin, Belfield, Dublin 4, Ireland
\label{dublin}      \and
School of Physics \& Astronomy, University of Southampton, University Road, Southampton SO17 1BJ, UK
\label{southamp}    \and
Departments of Physics and Astronomy, Le Conte Hall, University of California, Berkeley, CA 94720, USA
\label{ucb}      \and
   European Southern Observatory, Casilla 19001, Santiago 19, Chile
\label{esoc}      \and
   Advanced Concepts Team, European Space Agency, TEC-SF, ESTEC, Keplerlaan 1, NL-2201, AZ Noordwijk, The Netherlands
\label{actesa}      \and
   University of Exeter, Physics Building, Stocker Road, Exeter EX4 4QL, United Kingdom
\label{exeter}      \and
   Fakult\"at f\"ur Physik, Universit\"at Duisburg-Essen, Lotharstraße 1, 47057 Duisburg, Germany
\label{duisburg}      \and
   Instit\"ut f\"ur Astronomie und Astrophysik, Universit\"at T\"ubingen, Auf der Morgenstelle 10, 72076 T\"ubingen, Germany
\label{tuebingen}      \and
   Center for Space and Habitability, Universit\"at Bern, Gesellschaftsstrasse 6, 3012 Bern, Switzerland
\label{bern}      \and
   Astronomy Department, University of Michigan, Ann Arbor, MI 48109 USA
\label{umich}      \and
   Academia Sinica, Institute of Astronomy and Astrophysics, 11F Astronomy-Mathematics Building, NTU/AS campus, No. 1, Section 4, Roosevelt Rd., Taipei 10617, Taiwan
\label{sinica}      \and
   European Space Agency (ESA), ESA Office, Space Telescope Science Institute, 3700 San Martin Drive, Baltimore, MD 21218, USA
\label{esa}      \and
   Department of Earth \& Planetary Sciences, Johns Hopkins University, Baltimore, MD, USA
\label{jhueps}      \and
   Max Planck Institute for Astrophysics, Karl-Schwarzschild-Str. 1, 85741 Garching, Germany
\label{mpa}      \and
   Excellence Cluster ORIGINS, Boltzmannstraße 2, D-85748 Garching bei München, Germany
\label{origins}      \and 
    DOTA, ONERA, 13661 Salon cedex AIR, France
\label{onera}      \and
KRP Mechatec GmbH, Boltzmannstraße 2, 85748 Garching bei M\"unchen, Germany
\label{KRP}     \and
Universidad de Valparaíso, Instituto de Física y Astronomía (IFA), Avenida Gran Bretaña 1111, Casilla 5030, Valparaíso, Chile
\label{univalp}     \and
Núcleo Milenio de Formación Planetaria (NPF), Valparaíso, Chile
\label{nucleo} \and
Univ. Lyon, Univ. Lyon 1, ENS de Lyon, CNRS, Centre de Recherche Astrophysique de Lyon UMR5574, 69230, Saint-Genis-Laval, France
\label{cral} 
}

   \date{Received 6 February 2024; accepted 24 March 2024}

% \abstract{}{}{}{}{} 
% 5 {} token are mandatory
 
  \abstract
   {Since 2019, GRAVITY has provided direct observations of giant planets and brown dwarfs at separations of down to 95~mas from the host star. Some of these observations have provided the first direct confirmation of companions previously detected by indirect techniques (astrometry and radial velocities).}
   { We want to improve the observing strategy and data reduction in order to lower the inner working angle of GRAVITY in dual-field on-axis mode. We also want to determine the current limitations of the instrument when observing faint companions with separations in the 30 to 150~mas range. }
   { To improve the inner working angle, we propose a fiber off-pointing strategy during the observations to maximize the ratio of companion-light-to-star-light coupling in the science fiber. We also tested a lower-order model for speckles to decouple the companion light from the star light. We then evaluated the detection limits of GRAVITY using planet injection and retrieval in representative archival data. We compare our results to theoretical expectations. }
   {We validate our observing and data-reduction strategy with on-sky observations; first in the context of brown dwarf follow-up on the auxiliary telescopes with HD~984~B, and second with the first confirmation of a substellar candidate around the star Gaia DR3 2728129004119806464. 
   With synthetic companion injection, we demonstrate that the instrument can detect companions down to a contrast of $8\times 10^{-4}$  ($\Delta \mathrm{K}= 7.7$~mag) at a separation of 35~mas, and a contrast of $3\times 10^{-5}$  ($\Delta \mathrm{K}= 11$~mag) at 100 mas from a bright primary (K<6.5), for 30 min exposure time.}
   {With its inner working angle and astrometric precision, GRAVITY has a unique reach in direct observation parameter space. This study demonstrates the promising synergies between  GRAVITY and Gaia for the confirmation and characterization of substellar companions.}


   \maketitle
%
%________________________________________________________________
\nolinenumbers
\section{Introduction}
Thanks to progress in ground-based and space-based direct imaging instrumentation, we can now delve into the specific formation processes leading to substellar companions, such as massive planets and brown dwarfs. It has been proposed that they form by core accretion \citep{mizuno1980}, disk instability \citep{boss1997}, or collapse of the prestellar core \citep{bonnell2008}, and all these models come with different variations. It is still unclear as to which mechanism dominates in each type of object and at what distance. Extensive direct imaging surveys \citep[e.g.,][]{shine3,GPIsurvey19,stone2018} have inferred the occurrence rates of massive Jovian planets and brown dwarfs around stars of spectral types from B to M. For intermediate FGK stars, the findings of these surveys favor a dichotomy in the formation processes. The distribution of giant planets within 50~au is consistent with the predictions of the core-accretion model, and the giant planet and brown dwarf populations further out are consistent with the disk-instability pathway. Also, by fitting the orbit of a sample of a dozen substellar companions, \cite{bowler2020} revealed a difference in the eccentricity distribution of giant planets and brown dwarfs. These authors suggest that planets form in disks and brown dwarfs preferably by core collapse. The most promising way to enlarge the samples for testing the formation theories is to enable direct observations of fainter companions, and to reach the closer-in regions of the systems (below 20~au).

On the one hand, this goal of observing fainter companions at shorter separations triggers the development of faster and higher-order adaptive optics (AO)\citep{boccaletti2022,lozi2022,Gplus2022} and deconvolution techniques in high-contrast images (e.g., angular differential imaging: \cite{marois2006}, spectral differential imaging: \cite{racine1999}). So far, ground-based AO-assisted single telescopes and space instruments achieve contrasts down to a few $10^{-7}$ at 1 arcsec separation (ERIS: \cite{davies2023}, SPHERE: \cite{beuzit2019}, GPI:\cite{macintosh2014}, HiRISE: \cite{otten2021}, KPIC:\cite{jovanovic2019}, JWST: \cite{hinkley2022jwst} ). On the other hand, optical long-baseline interferometry with GRAVITY is emerging as a complementary technique because its specific deconvolution capability allows direct observations of planetary companions at separations of as small as 90\,mas.

GRAVITY is a second-generation K-band instrument and a two-in-one interferometric combiner in operation at the Very Large Telescope Interferometer (VLTI) since 2016 \citep{firstlightGRAV2017}. The fringe tracker (FT, \cite{lacour2019}) arm operates at 1~kHz on a bright target (K<10) in order to adjust the delay-lines position and correct for atmospheric turbulence. In parallel, the science (SC) arm can integrate up to 300 seconds, and thus allows for observations of objects as faint as K=19~mag \citep{deepgravity2022}. In addition, a metrology system measures the angular separation between the FT and the SC in real time. The ability to observe faint objects, together with the robust metrology link between FT and SC, launched optical interferometry into the field of direct imaging of exoplanets. The ExoGRAVITY large program has already provided direct observations of exoplanets orbiting at 3~au from their stars, at challenging separations of down to 95~mas and contrasts of a few $10^{-5}$ \citep{nowak2020,lacour2021,hinkley2022}. GRAVITY provides the relative astrometry with a precision down to 50~$\mu$as and a near-infrared K-band spectrometry at R$\sim$500 (medium-resolution mode) or R$\sim$4000 (high-resolution mode). These observations provide unprecedented constraints on the companion's orbit and allow determination of the object's surface temperature and atmospheric composition \citep{nowak2020b}. However, the field of view of the instrument is limited due to the single-mode nature of modern optical interferometry. For more than 50\% injection, it is about 65\,mas on the unit telescopes (UTs) and 290\,mas on the auxiliary telescopes (ATs). While this is an important drawback when performing blind searches, the situation has completely changed with the release of the Gaia space telescope catalog. Gaia's Non-Single-Star (NSS) two-body orbit catalog \citep{holl2022gaia, halbwachs2022gaia} published within DR3 \citep{gaiacollaboration2022gaia} contains astrometry-based orbital solutions for approximately \SI{450000}{} stars around each system's center of mass. Assuming that the orbital motion of the star is caused by the presence of a dark and unseen secondary body, the orbital solution constrains the on-sky position of the companion relative to the star with sufficient accuracy to position the single-mode fiber of GRAVITY. A subsequent detection of the companion with GRAVITY can confirm the candidate and provide the dynamical mass ---thanks to the astrometry--- and a direct measurement of its luminosity and spectrum. This synergy has been identified for a few years now, but there is currently no quantitative assessment of its potential. The actual inner working angle and contrast performance of GRAVITY are still undocumented, and their limitations are still unknown. These questions become even more pressing in the context of the ongoing instrumentation upgrade at the VLTI \citep{eisen2019}.

The goal of this paper is to demonstrate and quantify the potential of GRAVITY to provide direct confirmations of substellar candidates detected by Gaia absolute astrometry and to  understand its
limitations. In Sect.~\ref{sect:method}, we describe specific details of substellar companion observations and data reduction with GRAVITY. We outline a strategy to lower the inner working angle. In Sect.~\ref{sec:results}, we quantify and validate these strategies by observing the brown dwarf HD~984~B and by providing the first direct observation of a brown dwarf companion orbiting the star Gaia DR3 2728129004119806464 (hereafter referred to as Gaia~...6464).
Finally, in Sect.~\ref{sect:detectionlimit}, we determine the detection limits of GRAVITY by injection and retrieval of synthetic companions in archival ExoGRAVITY observations. We compare the results with expectations from the fundamental statistical noise. We conclude the paper with a summary and a discussion of the synergy with Gaia and other direct-imaging instruments.

\section{Method} \label{sect:method}
\subsection{The ExoGRAVITY method} \label{sect:exomet}
The ExoGRAVITY community developed an observation technique and a dedicated pipeline to enable direct observations of exoplanets and brown dwarfs with the GRAVITY instrument down to a few tens of mas close to bright nearby stars \citep{nowak2020,nowak2020b}.

\subsubsection{Observing technique} \label{sect:obstech}
The VLTI recombines either the four relocatable ATs, each of $D=$1.8~m in   diameter, or the four UTs, each of $D=8$~m in  diameter. The beam from each telescope travels through the VLTI tunnels and delay lines to reach GRAVITY. In the instrument, the two combiners (FT and SC) are fed by separate optical single-mode fibers. In the dual-field mode of the instrument, the fibers of the SC and of the FT can be positioned at different locations on the focal plane of each telescope \citep{pfuhl2012}. For separations of less than 0.7 arcsec, the field is separated by a 50/50 beamsplitter before injection into the FT and SC fibers \citep[Appendix A in][]{nowak2020}. In a typical ExoGRAVITY observation sequence, the FT remains centered on the host star and the SC alternates between long integrations centered on the companion and shorter integrations centered on the star (to avoid saturation). The shape of the fiber mode can be approximated by a Gaussian beam with a full width at half maximum (FWHM) of 65~mas on the UT (290~mas on the AT), which defines the field of view of the instrument. The flux injected into the SC from individual telescopes is recombined in integrated optics \citep{perraut2018}. The recombination method allows us to measure the total flux of each of the four telescopes, but also the coherent flux of each of the six baselines.

\begin{figure}
\centering
\includegraphics[width=\hsize]{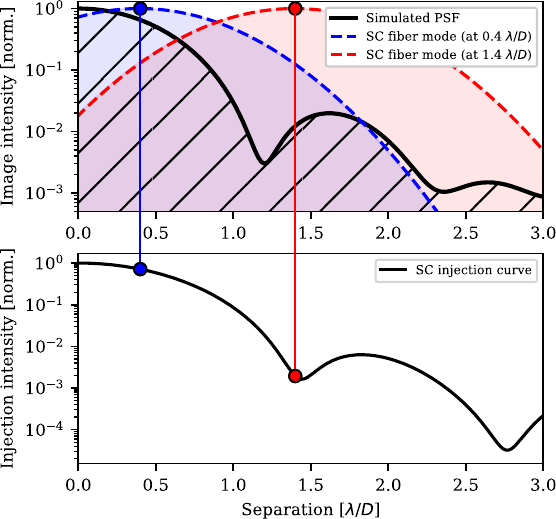}
   \caption{Injection into the SC arm at the focal plane of GRAVITY. Top: Comparison of the Gaussian single mode of the SC fiber with the diffraction-limited PSF at the image plane. This simulation is for 20\% bandwidth and the UT aperture. Bottom: Injection map given by the convolution of the PSF with the SC fiber mode.}
      \label{fig:injectionexplanation}
\end{figure}

Regarding the total flux (e.g., photometric flux of each telescope), the injected flux is the scalar product between the Gaussian mode of the fiber and the object point-spread function (PSF; Fig.~\ref{fig:injectionexplanation}). The dependence of the transmission on the distance $\mathbf{s}$ of the object from the center of the fiber is thus given by
\begin{align}\label{eq:transm}
    T(m,\mathbf{s})= \left| \iint  E(\mathbf{x},m)\, M(\mathbf{x}-\mathbf{s})\;\mathrm{d}\mathbf{x} \right|^2,
\end{align}
where $m$ is the telescope, $E$ the incident electric field, $M$ the fiber mode, and $\mathbf{x}=(x,y)$ the coordinates on the focal plane. In the following, we consider that $T$ includes the overall transmission from the atmosphere and the instrument; it therefore depends on the telescope, time, and wavelength. This way, the total flux $F_{\textrm{oncompanion}}$ injected into the SC for each of the four telescopes when the fiber is centered on the companion can be expressed as
\begin{align} \label{eq:exograv_tot}
    F_{\textrm{oncompanion}}(m,t&,\lambda) =\notag \\
    &F_s(\lambda)\,T(m,\boldsymbol{\Delta\alpha},t,\lambda) + F_c(\lambda)\,T(m,\mathbf{0},t,\lambda),
\end{align}
where $m$ is the telescope, $t$ the time, $\lambda$ the wavelength, $\boldsymbol{\Delta\alpha}=(\Delta \mathrm{RA}, \Delta \mathrm{Dec})$ the position of the companion relative to the star, and $F_s$ and $F_c$ are the total flux from the star and the companion, respectively.

The coherent flux, also called the complex visibility, encodes the amplitude and the phase of the interferometric fringes and provides the useful signal in GRAVITY. We can define the interferometric transmission $G$ as:
\begin{align}
    G(b,&\mathbf{s})=  \notag \\
    &\iint  E(\mathbf{x},m_1)\,M(\mathbf{x}-\mathbf{s})\;\mathrm{d}\mathbf{x} \cdot \iint  E^*(\mathbf{x},m_2) M^*(\mathbf{x}-\mathbf{s})\;\mathrm{d}\mathbf{x},
\end{align}
with the two telescopes $m_1$ and $m_2$ composing the baseline $b$. We can then write the complex visibility $V_{\textrm{oncompanion}}$ that the instrument measures when the SC fiber is located on the companion as
\begin{align} \label{eq:exograv1_cohe}
    V_{\textrm{oncompanion}}&(b,t,\lambda) =\notag \\
    &V_s(b,t,\lambda)\,G(b,\pmb{\Delta} \pmb{\alpha},t,\lambda) + V_c(b,t,\lambda)\,G(b,\mathbf{0},t,\lambda),
\end{align}
where $V_s$ and $V_c$ are the visibility of the star and the companion, respectively. Here and in the following, the visibility is understood as the complex coherent flux, which matches the definition of \cite{nowak2020}.

\begin{figure}
\centering
\includegraphics[width=\hsize]{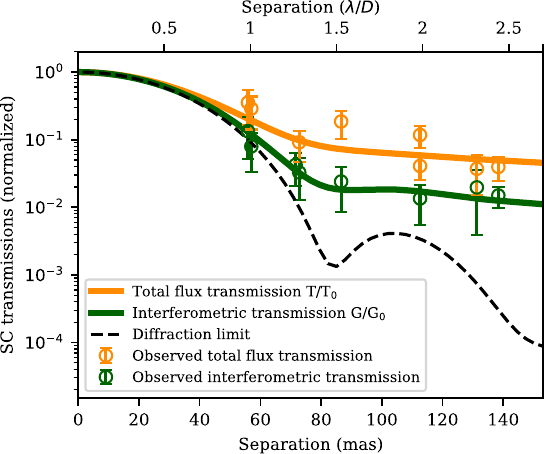}
  \caption{Total flux transmission ($T$) and interferometric transmission ($G$) as a function of the separation from the star on the UT. The points correspond to archival ExoGRAVITY observations around bright stars. The solid lines correspond to simulations matching the observations. The dashed line corresponds to a simulation of the transmission without atmospheric residuals but including a 20\% bandwidth and 10~mas rms tip-tilt jitter. }
     \label{fig:injectioncurve}
\end{figure}
The total transmission $T$ and the interferometric transmission $G$ include the flux losses due to the distance of the fiber from the object (star or companion). We estimated the dependence of the transmission on separation using archival observations from the ExoGRAVITY large program 1104.C-0651(A).  We selected observations with good atmospheric conditions. In the selected observations, the companion is fainter than contrasts of $10^{-4}$, and so the total flux $F_{\textrm{oncompanion}}$ and the coherent flux $V_{\textrm{oncompanion}}$ are largely dominated by the star contribution and we neglect the companion contribution. The GRAVITY pipeline \citep{lapeyrere2014} outputs an \textsc{astroreduced} file containing the total flux per telescope and per detector exposure (OI\_FLUX table) and the coherent flux per baseline and per detector exposure (OI\_VIS table). The ExoGRAVITY pipeline reads these files and normalizes the fluxes measured on-companion by the fluxes measured on-star. Therefore, the output immediately corresponds to the normalized injection curve.

Figure~\ref{fig:injectioncurve} displays the normalized coherent and total fluxes measured at different separations. These are based on archival observations from the ExoGRAVITY large program around bright stars from K=7.5~mag to K=3.5~mag ($\beta$ Pic, HD~206893, HD~17155, and CD-50~869). This dataset covers SC fiber positions from 55 to 140~mas, and atmosphere conditions from good to normal (seeing from 0.4 to 1.0 arcsec). To obtain continuous injection profiles, we ran AO simulations including single-mode fiber injection with HCIPy \citep{por2018hcipy}. We simulate an atmosphere following Kolmogorov turbulence and a low-order AO controlling 50 modes over an 8~m pupil to mimic the MACAO system at the Coudé focus of the UT \citep{arsenault2003}. We include 20\% bandwidth to account for the spectral range of GRAVITY from 1.95 to 2.4~$\mu$m. We also add realistic 10~mas rms tip-tilt jitter residuals from the VLTI tunnels \citep{anugu2018}. Finally, we adjust the AO loop gain and atmosphere parameters to match the observed total fluxes (coherent and total). The total flux transmission $T$ and the interferometric transmission $G$ are not at the diffraction limit level because of  atmospheric residuals not corrected for by the AO, turbulence in the VLTI tunnels, and (quasi-)static aberrations in the instrument. The simulations show that the average Strehl ratio in ExoGRAVITY observations is around 25\%, which is a realistic value. In the remainder of the paper, we use the continuous profiles to model the injection of coherent and total flux in GRAVITY.

\subsubsection{Unveiling the companion signal}\label{sect:datared}
The goal of the ExoGRAVITY reduction pipeline is to extract the companion astrometry and contrast spectrum from the measured complex visibilities $V_{\textrm{oncompanion}}(b,t,\lambda)$. In the following, "speckle" refers to the flux of the star that couples in the SC combiner while observing the companion. As shown in Eqs.~(\ref{eq:exograv_tot}) and (\ref{eq:exograv1_cohe}) and in Fig.~\ref{fig:injectioncurve}, this speckle light makes a contribution to the total flux, but also to the coherent flux. It is necessary to deconvolve the companion signal (coherent flux) from the coherent speckles. The companion is modeled as a point-source offset with respect to the host star. The complex visibility of the companion is:
\begin{align}\label{eq:planet}
    V_c(b,t,\lambda) = S_{c}(\lambda)\,e^{-i\frac{2\pi}{\lambda}\,[\mathbf{u}(t)\boldsymbol{\Delta\alpha}]},
\end{align}
where $S_{c}$ is the companion spectrum and $\mathbf{u}=(u,v)$ the coordinates of the array on the UV plane. In the ExoGRAVITY pipeline, the speckle term $V_sG$ of Eq.~\ref{eq:exograv1_cohe} is modeled as
\begin{align}
    V_s(b,t,\lambda)\,G(b,\boldsymbol{\Delta\alpha},t,\lambda) = P(b,t,\lambda)\,V_{\textrm{onstar}}(b,t,\lambda),
\end{align}
where $P$ is a complex polynomial that captures the spectral dependence of the coupling $G$ at the separation $\boldsymbol{\Delta\alpha}$, and $V_{\textrm{onstar}}$ is the visibility measured with the SC fiber centered on the host star:
\begin{align}
    V_{\textrm{onstar}} = J(b,t,\lambda)\; G(b,\mathbf{0},t,\lambda)\; S_{s}(\lambda).
\end{align}
 Here, $S_{s}$ is the star's spectrum, and $J$ is the function representing the drop in the visibility of the star if the star is resolved by the interferometer. In the following, we assume that the star is not resolved, and so $J=1$. Introducing the contrast spectrum 
\begin{align}
    C(\lambda)=S_{c}(\lambda)/S_{s}(\lambda),
\end{align}
we can rewrite Eq.~(\ref{eq:exograv1_cohe}) phase referenced on the star:
\begin{align} \label{eq:exograv2}
    V_{\textrm{oncompanion}}(b,t&,\lambda) = P(b,t,\lambda)V_{\textrm{onstar}} +C(\lambda)V_{\textrm{onstar}} e^{-i\frac{2\pi}{\lambda}\cdot[\mathbf{u}(t)\boldsymbol{\Delta\alpha}]}.
\end{align}
The first pass through the algorithm requires an assumption on $C(\lambda)$, and so, as a first guess, we assume that $C(\lambda)$ is a flat contrast spectrum. This first pass allows the recovery of the companion astrometry and the average flux ratio, and is the focus of this work. The second part of the pipeline (not described here) uses the astrometry to recover the companion contrast spectrum.

Equation~(\ref{eq:exograv2}) demonstrates the distinction between the speckle signal (first term) and the companion signal (second term). On the one hand, the speckle signal modulates at low spectral frequencies. On the other hand, the companion signal modulates at spectral frequencies that are determined by the projection of the companion separation onto the UV plane. This difference in the spectral oscillations allows the star light to be disentangled from the companion light (see Appendix~\ref{sec:ExoSIGNAL} for an example). This is the interferometric equivalent of the "spectral deconvolution" first introduced by \citet{2002ApJ...578..543S} for high-contrast imaging with  single telescopes.

\subsection{Aiming for a smaller inner working angle} \label{sec:pushing}

The previous section describes the standard ExoGRAVITY method. We now describe two modifications designed to improve the inner working angle.

\subsubsection{Off-pointing strategy} \label{sect:offpoint}
Figure~\ref{fig:injectioncurve} indicates that it should be possible to improve the flux ratio between the companion and the star by offsetting the position of the fiber when observing the companion. This adds a new degree of freedom in the observation: the offset $\boldsymbol{\delta}$ of the SC fiber with respect to the expected companion position. This way, Eq.~(\ref{eq:exograv_tot}) and (\ref{eq:exograv1_cohe}) can be rewritten as
\begin{align}
    F_{\textrm{oncompanion}}(m,t&,\lambda) =\notag \\
    &F_s(\lambda)\,
    T(m,t,\boldsymbol{\Delta\alpha}+\boldsymbol{\delta},\lambda) + F_c(\lambda)\,T(m,t,\boldsymbol{\delta},\lambda),
\end{align}
\begin{align}
V_{\textrm{oncompanion}}&(b,t,\lambda) =\notag \\
    &V_s(b,t,\lambda)\, G(b,t,\boldsymbol{\Delta\alpha}+\boldsymbol{\delta},\lambda) + V_c(b,t,\lambda)\,G(b,t,\boldsymbol{\delta},\lambda).
\end{align}
To reduce the star light without excessively reducing the coupling of the companion light, the offset $\boldsymbol{\delta}$ must be in the direction away from the star and of only a fraction of the PSF central lobe width. The off-pointing technique takes advantage of the sharp decrease in stellar transmission and the moderate decrease in companion transmission when the fiber is moved away from the star by a small offset $\boldsymbol{\delta}$.
\begin{figure}
\centering
\includegraphics[width=\hsize]{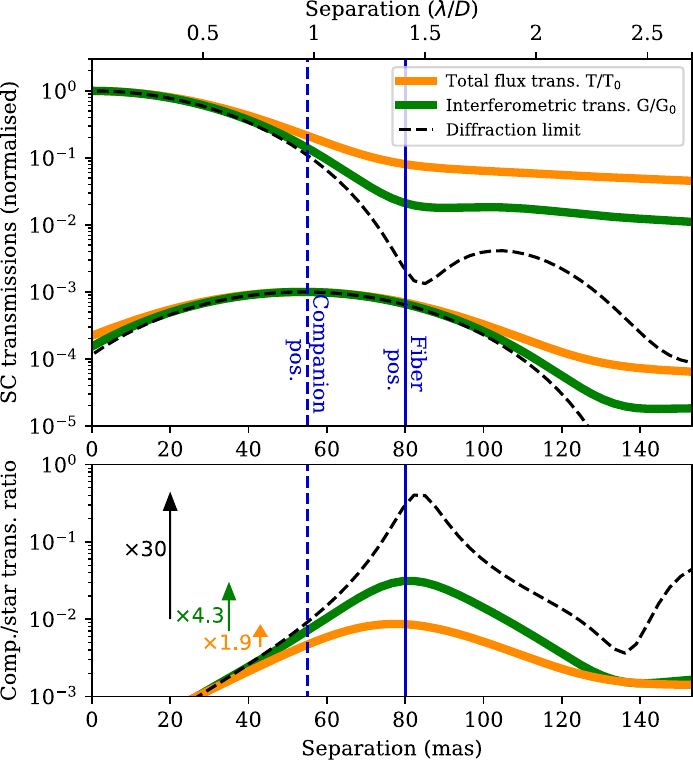}
  \caption{Quantification of the contrast improvement brought by the off-pointing strategy in the case of a companion at a  separation of 55~mas. Top: Dependence on separation of both transmission of the total flux ($T$) and interferometric transmission ($G$) for a star and a companion. The diffraction limit represents an ideal unaberrated case. Bottom: Companion-to-star transmission ratio with respect to separation. The arrows indicate the gain in contrast with a +25~mas SC fiber offset. }
     \label{fig:ttdh}
\end{figure}

\subsubsection{Order of the polynomial fit} \label{sect:polyspec}

Section~\ref{sect:datared} implies that the degree of the polynomial $P$ used to reject the speckles determines the inner working angle of the ExoGRAVITY technique. There is a trade-off between the quality of the speckle fit, and the self-subtraction of the planet signal. This is common to most deconvolution techniques.
By default, the ExoGRAVITY pipeline uses a fourth-degree polynomial for $P$. This has proven necessary to deal with the frequent fringe jumps of the fringe tracker. Such phase jumps cause a loss of visibility at the edges of the K-band visibility spectrum. These visibility losses are not necessarily of the same amplitude when observing on the star and on the planet, depending on the non-stationary quality of the fringe tracking. A fourth-degree polynomial is required to capture and remove these features. Consequently, most results published so far with the ExoGRAVITY pipeline have used a polynomial of between fourth and sixth order.

However, the situation recently changed with the commissioning of an improved version of the fringe tracker hardware and tracking algorithm in November 2022 \citep{Abuter+2016,nowak2024ft}. The new FT update significantly  reduces the occurrence of fringe jumps and thus relaxes the requirements on the degree of the polynomial. It is also interesting to explore the impact of this parameter because the planned upgrade of the AO  will stabilize the Strehl, and thus further reduce the occurrence of fringe jumps.

\subsection{Empirical detection limit}
\label{sec:injret}

The standard ExoGRAVITY pipeline lacks a method for determining robust detection limits. The injection of synthetic companions and their retrieval in the data is a classical approach to assessing the limit of direct imaging techniques in realistic conditions. The first step is to create a data set without a companion signal (if the original observation contains one). For this, we extract the companion astrometry $\Delta$RA,$\;\Delta$Dec and the contrast spectrum $C(\lambda)$ exposure by exposure thanks using the ExoGRAVITY pipeline (Sect.~\ref{sect:datared}). We then subtract the signal of this companion in the VISDATA table of the \textsc{astroreduced} files. We note that, unlike most other interferometric instruments, this operation is linear because GRAVITY operates with first-order estimators (complex coherent flux) instead of higher-order estimators (power spectrum and bispectrum). We then add synthetic companions at a given contrast $C$ and $\boldsymbol{\Delta\alpha}$ position to the VISDATA:
\begin{align} \label{synthplanet}
    V_{\textrm{syntheticcomp}}=C\,V_{\textrm{onstar}} \; e^{-i\frac{2\pi}{\lambda}\,[\mathbf{u}(t)\boldsymbol{\Delta\alpha}]} \; e^{i\varphi}.
\end{align}
The additional $\varphi$ phase term is due to the fact that the complex visibilities VISDATA are phased on the fiber position, and that the phase reference is the fringe-tracking phase, which is not necessarily zero. The term $\varphi$ writes: 
\begin{align} \label{eq:phasereference}
    \varphi = arg(\textrm{STAR\_REF})-&\textrm{PHASE\_REF} \\ \notag
    &+\textrm{PHASE\_MET}+\frac{2\pi}{\lambda}\textrm{DISP},
\end{align}
following the nomenclature of the GRAVITY pipeline user manual\footnote{\url{https://www.eso.org/sci/software/pipelines/index.html\#pipelines_table}}, where $\textrm{PHASE\_REF}$ is the phase of the fringe tracker, $\textrm{PHASE\_MET}$  the differential phase between the fiber coupler and telescope diodes, $\textrm{DISP}$ the fiber differential delay lines (FDDL) delay, and $\textrm{STAR\_REF}$ is the average of the two closest acquisitions with the SC on-star. In Eq.~(\ref{eq:phasereference}), we use a $\textrm{STAR\_REF}$ expression that has been previously rephrased with the metrology ($\textrm{PHASE\_MET}$) and FDDL delay ($\textrm{DISP}$). In this paper, we inject companions with a flat contrast spectrum for the sake of simplicity, and so $C$ is scalar. The modified VISDATA are then reduced by the ExoGRAVITY pipeline. 

We consider the detection successful if the companion is recovered less than 3~mas away from the injected position and at a contrast with less than 50\% relative error compared to the injected contrast. We consider that the detection limit is reached when fewer than 68\% (1$\sigma$) of the synthetic companions are successfully retrieved.

\section{Results} \label{sec:results}

\subsection{Expected performance improvement}

\subsubsection{Off-pointing technique}
We use the injection profiles shown in Fig.\ref{fig:injectioncurve} to quantify the improvement that can be achieved by the off-pointing technique. Figure~\ref{fig:ttdh} shows that an offset of the fiber position can result in a factor 4.3 improvement in the companion/star coherent flux ratio and a factor 1.9 improvement of the total flux ratio. In this example, the companion is at 55 mas from the host star, and the fiber is positioned at 80~mas (55~mas + 25~mas away from the star). The fiber offset from the companion results in an injection efficiency of 67\% (it would
be 100\% if the fiber were centered on the companion). This flux loss due to the offset must be compensated for a posteriori in order to recover the correct companion magnitude (\cite{wang2021}, Appendix A). We estimate that, with the current AO on the UT, this method can bring a contrast enhancement in coherent flux of up to a factor 6. In terms of the implementation, for companions with a separation of less than $55$~mas, the SC fiber should be placed at $+25$~mas from the companion, because a higher offset would result in more planet flux loss and possible errors in the astrometry due to aberrations in the fiber injection \citep{improvedastrom2021}. For companions with a separation of between $55~\text{mas}$ and $80$~mas, the SC fiber should be placed at 80~mas from the star. The offset method is not valid for companions with separations of greater than $80$~mas, and so the SC fiber should be placed on the companion  for these targets.

The improvement provided by the off-pointing strategy is currently severely limited by the AO performance. The dashed curve in Fig.~\ref{fig:ttdh} represents a realistic improved version of VLTI, with a high-order AO and control of the instrument aberrations. The expected injection profile was computed with the diffraction limit plus a 10 mas tip-tilt jitter, for instance resulting from the VLTI tunnels. The curve also considers a 20\% wavelength bandwidth. Under these conditions, off-pointing can bring a contrast enhancement of up to a factor 30. This result emphasizes the need for better AO and instrument aberration control, as will be implemented with GRAVITY+.

\subsubsection{Order of the polynomial fit} 

We used the planet injection and retrieval method described in Sect.~\ref{sec:injret} to define the most relevant polynomial degree when observing at short separation. We injected planets in the data set from the Gaia~...6464~B run on the UT with the SC fiber at 60~mas of the primary (described in detail in Sect.~\ref{sect:G272}). We then determined the best polynomial order to use in the reduction depending on the expected separation of the companion.

\begin{figure}
\centering
\includegraphics[width=\hsize]{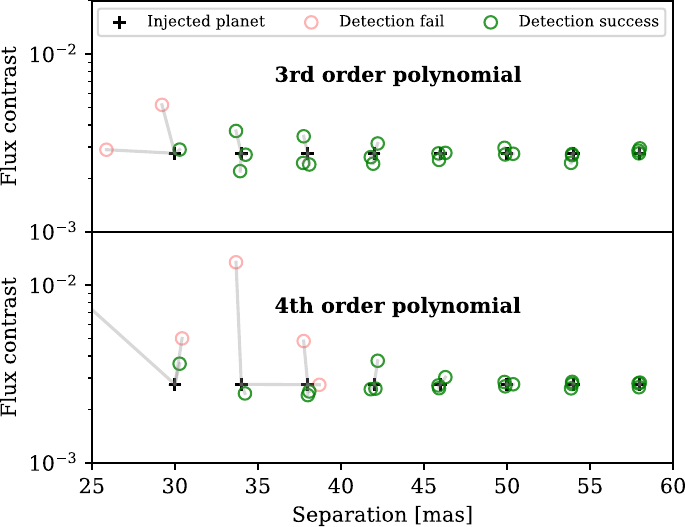}
  \caption{Planet retrieval at $2.7\times 10^{-3}$ raw contrast and separations from 30 to 90 mas. At each separation, three companions are injected at different positional angles to test the robustness of the retrieval. We compare the third-degree (top) and fourth-degree (bottom) polynomials for speckle modeling.}
     \label{fig:poly34}
\end{figure}

Figure \ref{fig:poly34} compares the performance of the third- and fourth-order polynomials for speckle modeling. The third-order polynomial leads to better planet retrieval below 45~mas separation, for both position and contrast. Beyond 45~mas, the fourth-order polynomial gives slightly better results for contrast retrieval. For completeness, we also ran tests with second-order polynomials. In this case, the retrieval gives erratic results. Thus, we still consider it safer to use the fourth order for separations of greater than 45~mas, as it gives more degrees of freedom for speckle modeling; nevertheless, we recommend using a third-order polynomial below 45~mas separation. This improvement brings the effective innermost working angle of GRAVITY to about 30~mas. 

\subsection{On-sky validation on the auxiliary telescopes}
\label{sect:validation}

\subsubsection{Observation and data reduction} 
The star HD 984 A is known to host a companion brown dwarf with a contrast at K-band of $3.7\times 10^{-3}$ \citep{meshkat2015,franson2022}. We used this binary system to validate the off-pointing technique described in Sect.~\ref{sect:offpoint}. The observation was performed as part of the program 0110.C-0182(A), on the ATs in astrometric configuration, and with the medium spectral resolution of GRAVITY. The off-pointing technique described for the UT is still valid on the AT, but since the ATs are smaller, it requires a $\times 4.4$ scaling on the angular separations. From orbit fits using previous observations \citep{whereistheplanet2021}, the companion was expected at a separation of 255~mas, with $\pm5$~mas uncertainty on the ($\Delta$RA,$\Delta$Dec). During the observation, the SC fiber was alternately positioned on the predicted position of the companion and 100~mas further away from the star (+0.4~$\lambda/D$), as summarized in Table~\ref{tab:logobsHD984}. We chose a detector integration time (DIT) of 30s, we collected a number of acquisition (NDIT) equal to eight in each of the three files (NEXP), and this for each position of the SC. We reduced the data using the ExoGRAVITY pipeline. The data taken with and without the off-pointing technique were reduced separately to compare the results. After reduction, it appeared that the companion was 9~mas away from the expected position. As this corresponds to only 3\% of the fiber field of view on the AT, it did not entail flux losses and had no negative effect on our test.

\begin{table}[t]
\centering                          % used for centering table
\renewcommand{\arraystretch}{1.1}
\caption{Log for the GRAVITY observations of HD 984 AB on the AT. }
\begin{tabular}{c c c c}        % centered columns (4 columns)
\hline
\multicolumn{4}{c}{\textbf{ Date: 2022-10-24}} \\
Observing time & Airmass & $\tau_0$\tablefootmark{a} & Seeing\\
01:12:17/01:57:20 & 1.07-1.14 & 2.5-4.5 ms & 0.43-0.66" \\
\hline
\hline
Target & $\Delta$RA/$\Delta$Dec\tablefootmark{b} & \multicolumn{2}{c}{NEXP/NDIT/DIT} \\
HD 984 A & 0/0 mas & \multicolumn{2}{c}{2/8/10 s} \\
HD 984 B & 162/197 mas & \multicolumn{2}{c}{3/8/30 s} \\
HD 984 B & 224/273 mas & \multicolumn{2}{c}{3/8/30 s} \\
\hline 
\end{tabular}
\tablefoot{\tablefoottext{a}{Atmosphere coherence time}.\tablefoottext{b}{SC fiber position relative to the star.}}
\label{tab:logobsHD984}
\end{table}

\subsubsection{Contrast improvement} 
\begin{figure}
\begin{subfigure}{.23\textwidth}
  \centering
  \includegraphics[width=0.93\linewidth]{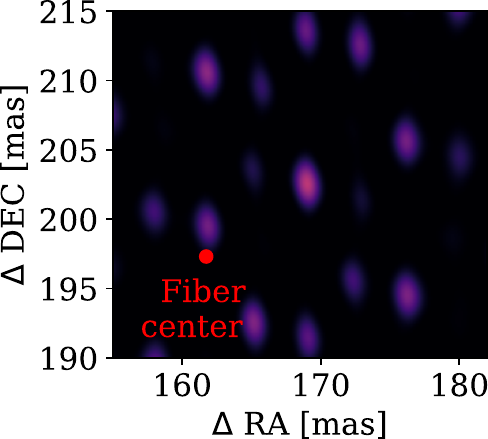}
  \caption{On-companion}
  \label{fig:sfig1}
\end{subfigure}%
\hspace{0.01\textwidth}
\begin{subfigure}{.23\textwidth}
  \centering
  \includegraphics[width=1.\linewidth]{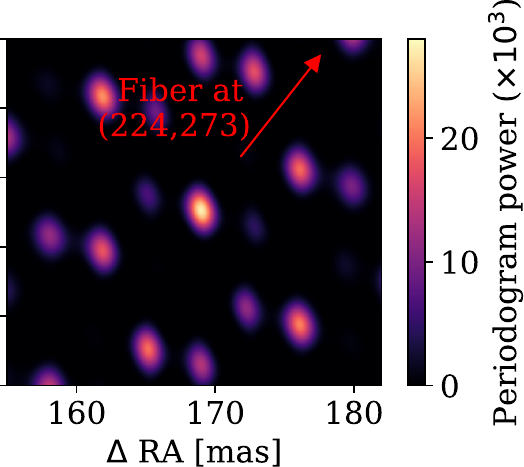}
  \hspace{-0.1\textwidth}
  \caption{Off-companion}
  \label{fig:sfig2}
\end{subfigure}
\caption{Periodogram resulting from the astrometry fit of the GRAVITY data on HD~984 obtained by pointing the fiber at the expected position of the companion (left) and using the off-pointing technique (right). The spot of higher periodogram power corresponds to the detection of HD 984 B.}
\label{fig:periodHD984}
\end{figure}

Applying the offset on the fiber position, the coherent flux injected into the SC is reduced by a factor 6 and the total flux is reduced by a  factor 3.4 on average. The slightly better improvement compared to expectations (factor 4 in coherent flux) is because the NAOMI AO on the ATs \citep{woillez2019} are closer to the diffraction limit than the UT with MACAO. The off-companion observations provide a significantly better detection than the on-companion, with about twice the periodogram power, from $1.4\times10^4$ to $2.7\times 10^4$, as shown in Fig.~\ref{fig:periodHD984}. 

Relative astrometry is improved by the off-pointing technique. With the off-pointing observation, the uncertainty on the $\Delta$RA is reduced by 35\% and the uncertainty on $\Delta$Dec is reduced by 8\% compared to the classical on-companion pointing.

\begin{figure}
\centering
\includegraphics[width=1\hsize]{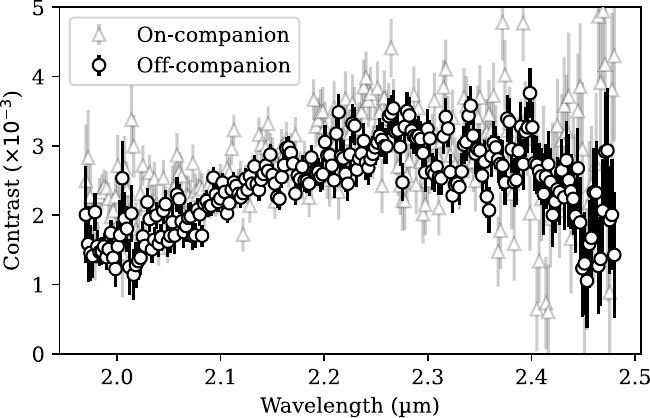}
  \caption{Contrast spectrum for HD 984 B from our observations with (black) and without (gray) the off-pointing technique.}
     \label{fig:spectreHD984}
\end{figure}

We also ran the second part of the pipeline to obtain the contrast spectrum. The spectrum quality is expected to improve thanks to the off-pointing. On the one hand, with the reduction of the total flux, we expect the amplitude of the photon noise from star speckles to be reduced by a factor of 1.8. On the other hand, only 72\% of the companion flux is injected due to the off-pointing. Overall, we expect an improvement of 30\% of the spectrum signal-to-noise ratio (S/N). The resulting contrast spectra are shown in Fig.~\ref{fig:spectreHD984}. The total S/N can be calculated as:
\begin{align}
    \text{S/N} = \sqrt{C^\intercal .COV^{-1}.C,}
\end{align}
where $C$ is the contrast spectrum, $C^\intercal$ its vector transpose, and $COV$ is the covariance matrix of the spectrum. The S/N is equal to 120 for the on-companion observations and is equal to 164 (+37\%) for the off-companion data. This is fully in line with expectations.

\subsubsection{Discussion}
The efficiency of the interferometric deconvolution depends on the length of the baselines and the separation. In other words, a companion at $1~\lambda/D = 252~$mas separation on the ATs is easier to disentangle from the speckles than a companion at $1~\lambda/D = 57~$mas on the UTs. Nevertheless, this observation of HD 984 B with the AT demonstrates that a $+0.4~\lambda/D$ offset brings a significant reduction of the coherent and total stellar flux, a better detection, and a higher S/N spectrum. It also shows that the technique does not introduce significant astrometric errors.

\subsection{Application to  Gaia~...6464~B on the unit telescopes} \label{sect:G272}

We compiled a list of objects from the Gaia NSS catalog with candidate substellar companions that are accessible for direct confirmation with GRAVITY \citep{winterhalder2024}. As a proof of concept, we observed the candidate Gaia~...6464 with the UTs during a technical time request for the science verification of a GRAVITY FT upgrade (60.A-9102). 

\subsubsection{Observations and data reduction} \label{sec:obs}
\begin{figure}
\centering
\includegraphics[width=0.9\hsize]{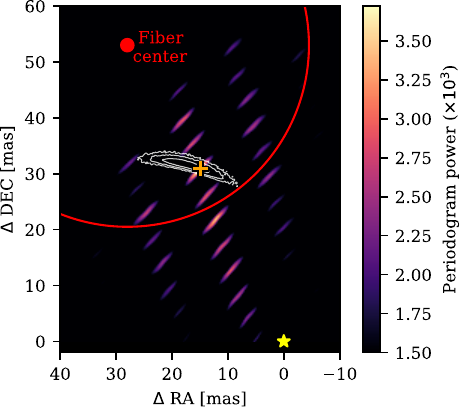}
  \caption{Periodogram from the astrometry fit of the observations of Gaia~...6464~B. The best-fit position of the Gaia~...6464~B according to the ExoGRAVITY pipeline is indicated by the orange plus symbol. The position of the primary star is marked by the yellow symbol. The white contour lines are the Gaia-based position probability density of the dark companion (from inside to outside: \SI{68}{}, \SI{95}{}, and \SI{99.7}{\%} of the total set of position predictions). The position of the SC fiber in the field is given by the red dot, while the fiber field of view (50\% flux injection limit) is shown by the red circle. }\label{fig:periodo_g272}
\end{figure}

We predict the position of Gaia~...6464~B  from the NSS catalog orbital solution, assuming that the companion does not contribute to the flux observed in the G-band and that the companion mass is the lower estimate listed in the Gaia DR3 \texttt{binary\_masses} table. We use a randomization procedure to obtain the projected position probability shown in Fig.~\ref{fig:periodo_g272}. The companion position is predicted with $\pm$5~mas uncertainty on separation, and $\pm 8^\circ$ uncertainty on positional angle at 1$\sigma$.  Given the short predicted separation for the companion (close to 35~mas), we pointed the science fiber 25~mas away from the predicted position to reduce the host star flux injection, as described in Sect.~\ref{sect:offpoint}. GRAVITY was set to medium spectral resolution and we used the dual-field on-axis mode. A detailed summary of the observing conditions and exposure time settings can be found in Table~\ref{table:obs}.

We applied the standard ExoGRAVITY pipeline, except that we reduced the polynomial order to 3 for the speckle fit, as described in Sect.~\ref{sect:polyspec}. Figure~\ref{fig:periodo_g272} shows the resulting periodogram. Gaia~...6464~B is detected at a separation of 34 mas and 2.5~mas from the predicted position. The contrast at K-band, corrected for fiber injection loss, is $(3.1\pm 0.5)\times 10^{-3}$. We confirmed the detection by injecting and recovering synthetic companions in the data at similar contrasts, as described in Sect.~\ref{sec:injret}. All the injected companions were correctly retrieved by the pipeline.

\begin{table}
\centering                          % used for centering table
\renewcommand{\arraystretch}{1.1}
\caption{Log for the GRAVITY observations on the unit telescopes.}
\begin{tabular}{c c c c}        % centered columns (4 columns)
\hline
\multicolumn{4}{c}{\textbf{ Date: 2022-11-09}} \\
Observing time & Airmass & $\tau_0$ & Seeing\\
01:51:09 / 02:57:59 & 1.54-2.16 & 3.5-9.6 ms & 0.35-0.88" \\
\hline
\hline
Target & $\Delta$RA/$\Delta$Dec\tablefootmark{a} & \multicolumn{2}{c}{NEXP/NDIT/DIT} \\
Gaia~...6464~A & 0/0 mas & \multicolumn{2}{c}{4/12/10 s} \\
Gaia~...6464~B & 28/53 mas & \multicolumn{2}{c}{16/4/30 s} \\
\hline 
\end{tabular}
\tablefoot{\tablefootmark{a}{SC fiber position relative to the star.}}             % title of Table
\label{table:obs}      % is used to refer this table in the text
\end{table}

\subsubsection{Dynamical mass determination}
The use of the fiber off-pointing and the lower-order polynomial for speckle fitting allowed a direct observation of Gaia~...6464~B at the innermost separations possible with the ExoGRAVITY technique. This detection can be used to infer the dynamical masses of the companion and primary, and to further constrain the orbital solution presented in the Gaia NSS catalog. To this end, we require two pieces of information from the GRAVITY direct imaging: confirmation of the dark companion hypothesis and the precise relative astrometry.

We used a Markov Chain Monte Carlo (MCMC) framework based on \cite{emcee_2013} to combine Gaia and GRAVITY data. It should be noted that Gaia NSS data must be handled with care when used in combination with MCMC methods, especially for low-eccentricity orbits \citep{Babusiaux_Gaia_DR3_catalogue_validation}. We used BINARYS \citep{Leclerc_BINARYS} to confirm that our MCMC analysis of Gaia~...6464~B is well-behaved and gives results similar to the those obtained with the local linear approximation technique. 

The direct observation provides the companion-to-star flux ratio, which is crucial for mass determination. We compared the mass results under two different assumptions: either the photocenter observed by Gaia is fully coincident with the primary position (dark-companion assumption), or the companion-to-primary flux ratio in the Gaia band in the visible is the same as the flux ratio measured at K band (faint-companion assumption). We find a companion mass that is \SI{2}{M_{\mathrm{Jup}}} higher under the faint-companion assumption and comparable error bars of the order of \SI{3}{M_{\mathrm{Jup}}} under the two assumptions. As the two hypotheses give similar results, and given that the flux ratio in the G band is expected to be lower than in the K band, the following orbital fit and mass determination remain under the dark-companion assumption.

\begin{figure}
    \centering
    \includegraphics[width=0.98\columnwidth]{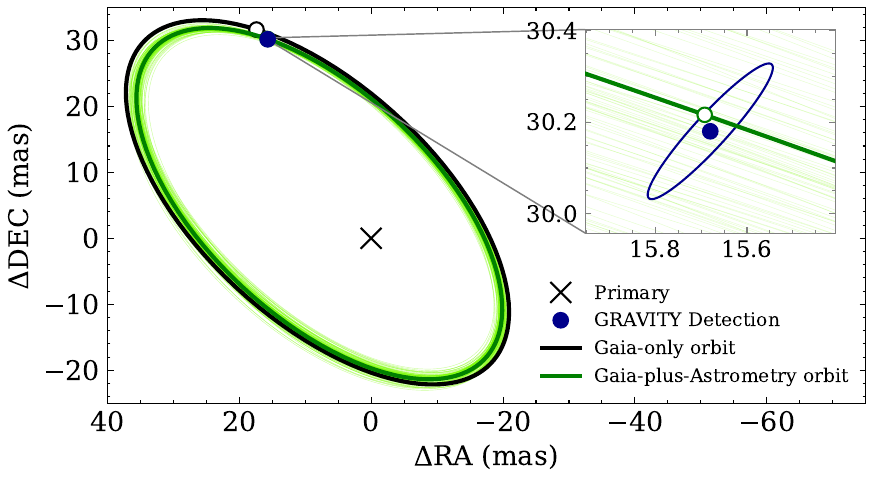}
    \caption{Orbit of the companion Gaia~...6464~B. The original Gaia orbit is shown in black. The refined orbit resulting from the combination of the Gaia solution and the astrometric position measurement provided by GRAVITY is shown in dark green. The light green shows a random subset of the sampled solutions probed during the MCMC run. The blue dot indicates the GRAVITY measurement with the ellipse describing the associated uncertainty. On the main plot, the black circle shows the position of the companion at the time of the observation as predicted by the Gaia-only orbital solution. On the inner plot, the green circle shows the position of the companion from the refined orbital solution including the GRAVITY observation.}\label{fig:orbit_plot}
\end{figure}

The MCMC-based combination of observations from Gaia and GRAVITY results in an updated set of posterior distributions. In a Bayesian sense, these updated posteriors correspond to a more accurate description of the system, unless the Gaia and GRAVITY data conflict in some significant way. The initial and updated orbital solutions projected onto the sky plane are shown in Fig.~\ref{fig:orbit_plot}. The Gaia-only solution is derived from the NSS catalog, and uses the primary mass derived by isochrone fitting (listed in the Gaia DR3 \texttt{binary\_masses} table). The Gaia-only orbit also assumes the lower-limit mass ratio between companion and host. As expected, the Gaia-only orbit and the Gaia+GRAVITY orbit are in good agreement. While the Gaia astrometry alone cannot constrain the individual masses, it is striking that a single observation with GRAVITY is sufficient to narrowly constrain the mass of the companion and the primary. We obtain $M_1$\,=\,$0.53_{-0.02}^{+0.02}$\,\SI{}{M_\odot} and $M_2$\,=\,$78.34_{-2.50}^{+2.62}$\,\SI{}{M_{\mathrm{Jup}}}. The detected companion sits at the upper limit of the conventional mass range of brown dwarfs (\SI{13}{} to \SI{80}{M_{\mathrm{Jup}}}). A detailed description of the method, along with the study of this companion and other targets are presented in \cite{winterhalder2024}.

\section{Detection limits} \label{sect:detectionlimit}

Evaluation of the instrumental limits of detection is necessary in order to properly use nondetections in a statistically significant manner. Such an evaluation also allows the instrument to be placed in the larger instrumental landscape. This information is not yet available in the literature for GRAVITY. The goal of this section is to provide quantitative numbers on this question. Furthermore, this work allows us to comment on the possible nature of the noise process that limits the achievable contrast.

\subsection{Contrast curves}\label{sect:contrastcurv}
We used the injection and retrieval technique described in Sect.~\ref{sec:injret}. The data set consists of five archival ExoGRAVITY observations on the UTs (details in Appendix~\ref{sec:ExoObs}) with the position of the SC fiber ranging from 54 to 136 mas. Archival observations are selected based on the criteria of having a bright primary (K<6.5) and good atmospheric conditions (seeing<0.85~arcsec). The integration time for all observations is $5\times32\times10$ seconds (27 minutes) spread over one hour to take advantage of the rotation of the UV plane. We simulated the fiber off-pointing technique below 80~mas separation (Sect.~\ref{sect:offpoint}) by injecting the synthetic companion not at the fiber separation but closer to the host star. The retrieval process is computationally intensive, and so we limited the number of companions injected to five per separation and contrast. The five different injections were performed at different positional angles to shuffle the possible effects of individual interferometric speckles in the observations.  The separation of the injected companions is chosen in a range of maximum $\pm 25$~mas around the position of the SC fiber (see Table~\ref{tab:exoobs}). The ability to detect companions is significantly affected by the positional angle of the star companion with respect to the UV plane. Therefore, in the following study, we chose the positional angles for planet injection based on the orientation of the VLTI baselines and not on the positional angle of the SC fiber. We tested companion retrieval at the most optimal angle (i.e., parallel to the longest baselines on the UV plane) and at the least optimal angle (i.e., perpendicular to the longest baselines). After the companion injection, we ran the ExoGRAVITY pipeline to retrieve the companion signal. We average the signal from the 32 NDITs of individual exposures before reduction (fast mode of the ExoGRAVITY pipeline) in order to perform hundreds of injections and retrievals within a reasonable time frame.

\begin{figure}
\centering
\begin{subfigure}{\columnwidth}
\includegraphics[width=\hsize]{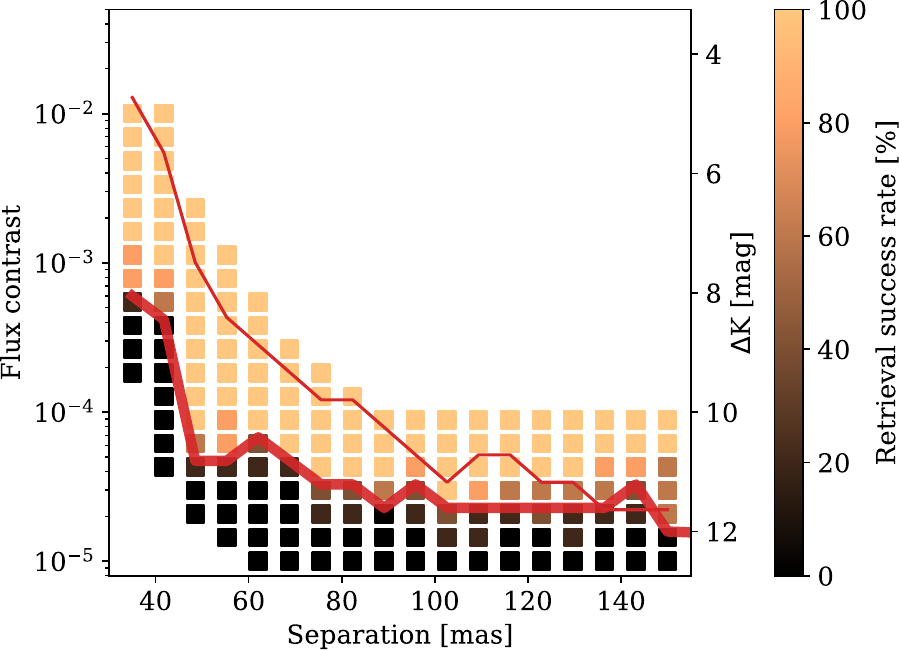}
  \caption{Best UV plane orientation}\label{fig:deteclimit_goodpa}
\end{subfigure}
\begin{subfigure}{\columnwidth}
\includegraphics[width=\hsize]{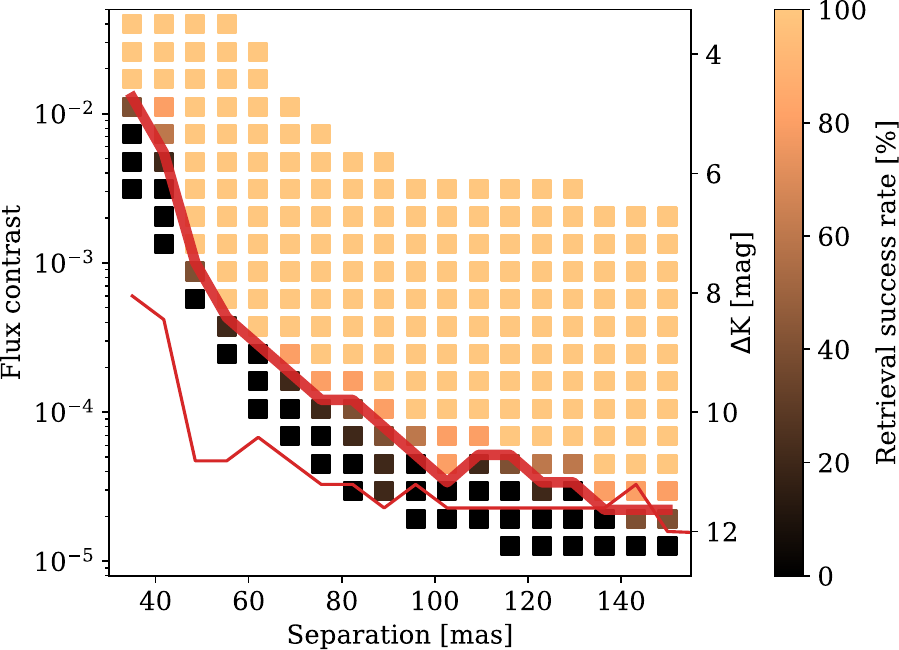}
\caption{Worst UV plane orientation}\label{fig:deteclimit_badpa}
\end{subfigure}
  \caption{Rates for successful retrieval of injected companions in ExoGRAVITY observations with 10s DIT and 27min exposure time on the UT. (a) Retrieval when the star--companion couple is oriented in the direction of the longest VLTI baselines. (b) Retrieval when the star--companion couple is oriented perpendicular to the longest VLTI baselines. The red curves are the 1$\sigma$ contrast limits for the best and worst UV plane orientation, shown as as thick or thin lines alternatively depending on the background plot. }\label{fig:deteclimit}
\end{figure} %$contrast= 2\times 10^6 (separation [mas]) ^{-6} + 4\times 10^{-5}$

The results are shown in Fig.~\ref{fig:deteclimit}. The contrast curve is obtained at 1$\sigma$ when fewer than four out of the five companions are retrieved at the separation and contrast investigated. We cannot significantly determine 2$\sigma$ or 3$\sigma$ limits because of the limited number of companions injected. When the positional angle is parallel to the longest VLTI baselines, the sudden loss of sensitivity between 40 and 50 mas is due to the subtraction of the companion signal by the polynomial (Sect.~\ref{sect:polyspec}). At separations of greater than 90~mas, the detection limit reaches $3\times 10^{-5}$ contrast ($\Delta \mathrm{K}= 11.3$~mag). When the positional angle is perpendicular to the longest VLTI baselines, the contrast limit below 70~mas is ten times shallower than in the best position-angle case. The relative orientation must be considered when planning an observation. Beyond 130~mas separations, the contrast limit reaches $3\times 10^{-5}$ and the UV plane orientation no longer has any influence. 

Finally, we explored the performance at larger separations. To this end, we performed additional injection and retrieval tests at 320~mas separation on AF Leporis b observations. The detection limit is at a contrast of $5.1\times 10^{-6}$ ($\Delta \mathrm{K}= 13.2$~mag).

\subsection{Empirical analysis of limitations} \label{sec:limitations}
Detection on GRAVITY can be limited by either correlated noise or statistical noise in the measured visibilities. Correlated noise can be caused by stellar speckles that are not correctly fitted by the polynomial modeling. White noise is due to photon noise and detector noise.

We investigated the detection limit for different observing times using archive observations on the UT on HD 206893 (with the fiber at 111~mas separation) and $\beta$ Pictoris (with the fiber at 92~mas separation). Both data sets have excellent atmospheric conditions and are bright enough for the AO to operate at the nominal regime. The HD~206893 data set has an exposure time of $27\times 32 \times 10$~s (2.4~h) and spans over 3.5~h. The $\beta$ Pictoris data set has an exposure time of $13\times 32 \times 10$~s (1.2~h) and spans over 2.2~h. We select successive exposures in the data set to mimic a shorter observation time. 

\begin{figure}[t]
\centering
\includegraphics[width=0.95\hsize]{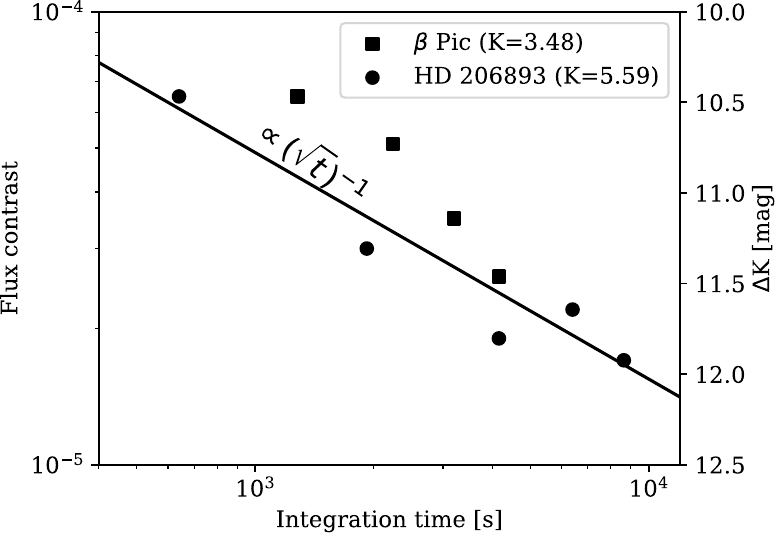}
  \caption{Limit contrasts at 1$\sigma$ with respect to the total exposure time (NEXP$\times$NDIT$\times$DIT). The solid line corresponds to the inverse square root of the integration time, with a proportionality constant of 1.5$\times$10$^{-3}$ in flux contrast.}\label{fig:deteclimitVSobstime}
\end{figure}

The results are summarized in Fig.~\ref{fig:deteclimitVSobstime}. The evolution of the limit contrast roughly follows the inverse square root of the integration time. However, $\beta$ Pictoris is seven times brighter in K-band than HD~206893. If limited by the photon noise of stellar flux leaking into the SC fiber, we expect the contrast to be $\sqrt{7}\approx 2.6$ deeper on $\beta$ Pictoris than on HD~206893 at the same integration time. Instead, we observe that the detection limit follows a similar $\sqrt{t}^{-1}$ trend for both stars. This indicates that the detection is not limited by photon noise, but by other features that are still averaging over the integration time. The situation is similar in classical imaging where the detection limit evolves with the characteristic lifetime of the speckles in the field. From an operational point of view, we can conclude that extending the integration time up to 90~min is a way to push the contrast limit down to $2\times 10^{-5}$ ($\Delta \mathrm{K}= 11.7$~mag) at 100~mas.

Overall, GRAVITY seems to be limited by systematic uncertainties that scale with the speckle flux. This explains why the off-pointing technique described in Sect.~\ref{sect:offpoint} has a significant impact on the detection limit, as would any further reduction of the stellar leak.

\subsection{Comparison with the theoretical limit} \label{seq:theoreticallimit}
The detection limits can also be determined analytically. Considering the photon noise from the star flux leaking in the fiber at the companion position and the readout noise of the SC camera, we can derive a S/N using
\begin{align} \label{eq:snr}
    &S/N = \notag\\
    &\frac{ \eta \; \sum_{b,\lambda}|G(b,\lambda)\,V_c(b,\lambda)|\,\mathrm{DIT}\,\mathrm{NDIT}_{\textrm{tot}}}{\sqrt{ \sum_{m,\lambda} T(m,\lambda)\,F_s(m,\lambda)\,\mathrm{DIT}\,\mathrm{NDIT}_{\textrm{tot}} + RON^2\,N_{\textrm{pixels}}\,\mathrm{NDIT}_{\textrm{tot}}}},
\end{align}
where $\eta{}$ is the fraction of companion signal remaining after subtracting the speckles polynomial, NDIT$_{\textrm{tot}}$ is the total number of integrations (NEXP$\times $NDIT), $RON$ is the read-out noise of the SC camera (Teledyne H2RG), and $N_{\textrm{pixels}}$ is the number of pixels used on the detector. The other parameters are described in Eqs.~(\ref{eq:exograv_tot}) and (\ref{eq:exograv1_cohe}).  

\begin{figure}
    \centering
    \includegraphics[width=1\hsize]{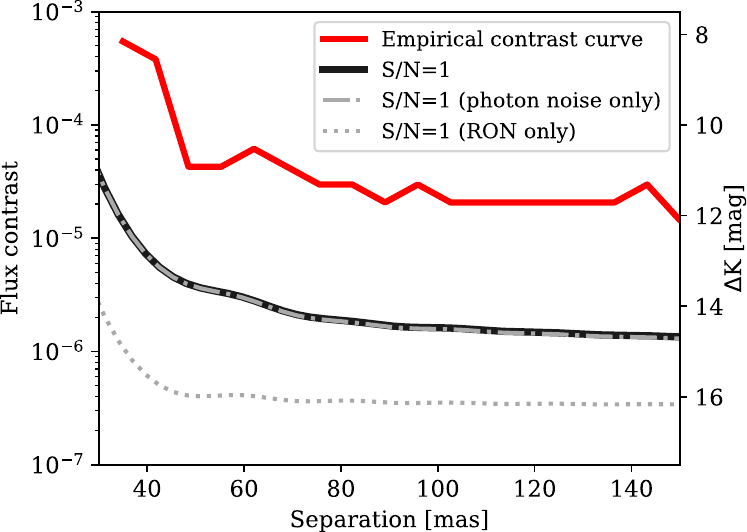}
    \caption{Theoretical contrast curves for S/N=1 (solid black) compared to the contrast curve derived from the injection and retrieval at a favorable positional angle (red). An indicative contrast curve is given with only the camera readout noise as a noise source (dotted black) and only the photon noise (dashed-dotted black). The theoretical contrast curves are computed for a host star of K-band magnitude 5.6 similar to HD 206893 used in the injection--retrieval data set.}\label{fig:theoreticalcurve}
\end{figure}

From Eq.~(\ref{eq:snr}) and parameter values listed in Table~\ref{tab:theoretical}, we can derive a limiting planet-to-star contrast for a given $S/N$. The results are shown in Fig.~\ref{fig:theoreticalcurve}. The photon noise from the star largely dominates the statistical noise. The detection limit at $S/N=1$ allows direct comparison with the 1$\sigma$ empirical contrast curve. This comparison shows that the theoretical detection limits are a factor 12 lower than the empirical detection limits determined from injection and retrieval. We show in Sect.~\ref{sec:limitations} that photon noise is not the limitation in actual observations at 100~mas separation. The findings shown in this section confirm that result.

\begin{table}
\centering
\caption{Parameters for the computation of the theoretical contrast curve. }
\begin{tabular}{c c } 
 \hline
Parameter & Value \\
\hline
DIT & 10 s\\
NDIT$_{\textrm{tot}}$ & 160 \\
RON & 9 photons/DIT \\
$N_{\textrm{pixels}}$\tablefootmark{a} & 11184 \\
\hline
Fluxes &   \\
\hline
$T_0$\tablefootmark{b} & 1 \% \\
$F_s$\tablefootmark{c} & $7\times 10^8$ photons/s \\
$V_c$\tablefootmark{d} & 0.8 $\times F_c$
\end{tabular}
\tablefoot{\tablefoottext{a}{There are 24 ABCD outputs on the detector, 233 pixels in the spectral direction, and the spectrum is two pixels wide.}\tablefoottext{b}{Total transmission of the VLTI and GRAVITY when the SC fiber is at 0~mas separation.} \tablefoottext{c}{Flux of HD 206893 for a single telescope. Computed from the star magnitude in K-band and the UT collecting surface. } \tablefoottext{d}{We estimate the instrumental visibility to be 0.8.} }
\label{tab:theoretical}
\end{table}

\section{Discussion}
\subsection{Summary}

\begin{figure*}[t]
\centering
\includegraphics[width=\hsize]{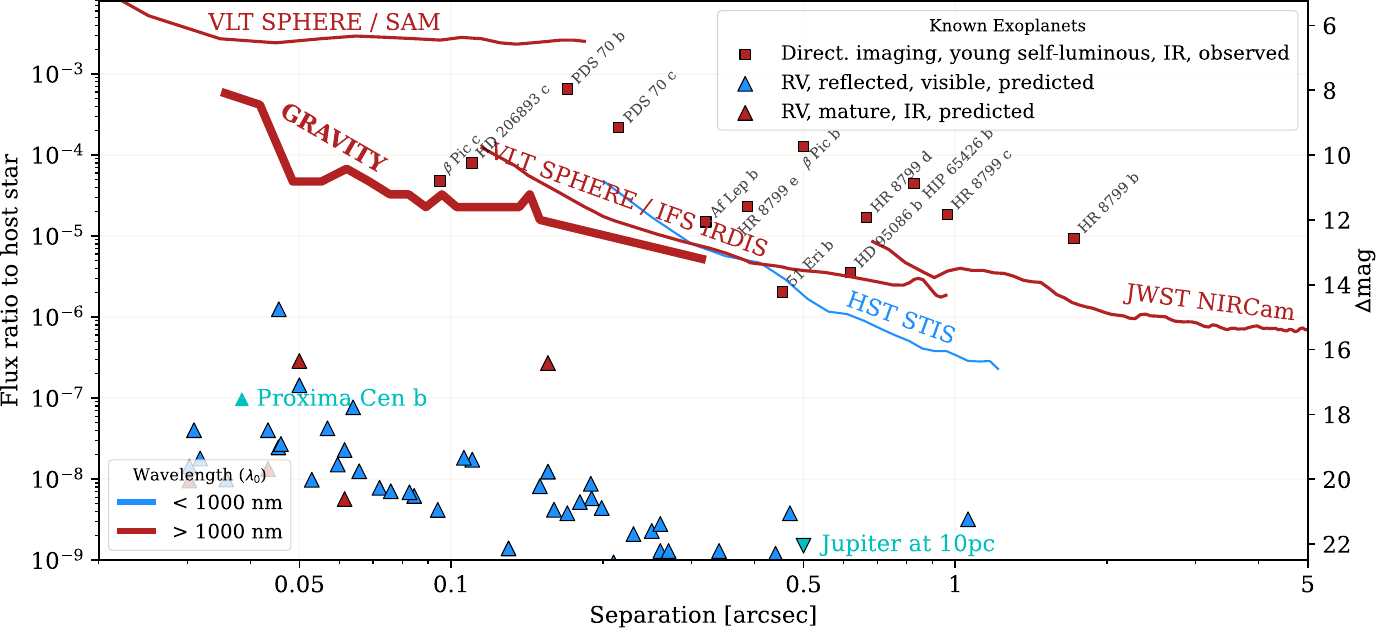}
  \caption[frog]{Detection limits of some representative instruments dedicated to direct observations of exoplanets. GRAVITY detection limit corresponds to the inner part of the on-axis mode of the instrument only. It is limited to 27 min exposure time and companions oriented parallel to the longest baselines of the VLTI. Plot adapted from Dr. Bailey script available at \url{https://github.com/nasavbailey/DI-flux-ratio-plot}. The HST/STIS curve is from \cite{ren2017}. VLT SPHERE/IFS IRDIS is from \cite{langlois2021}. VLT SPHERE/SAM is from \cite{stolker2023}. VLT JWST/NIRCam is from \cite{carter2023}. (Blue triangles) Estimated reflected flux in the visible for exoplanets observed using the radial velocity technique. The estimation follows a Lambertian model with radii fixed at 1~R$_{Jup}$ and a geometric albedo of 0.5. (Red triangle) Estimated infrared flux for mature exoplanets observed using the radial velocity technique. Computed from equilibrium temperature estimates and planet radii fixed at 1~R$_{Jup}$. All visible and infrared estimated fluxes are based on \cite{traub2010}. }\label{fig:DI_curves}
\end{figure*}

In this paper, we detail a proposed observing strategy and adjustment of the data reduction pipeline required to enable exoplanet observations at shorter separations with GRAVITY. We validated the proposed strategy and demonstrated the Gaia--GRAVITY synergy with the detection and mass determination of the short-separation companion Gaia~...6464~B. We finally explored the actual capabilities of GRAVITY by deriving realistic contrast curves. The most important results of this work can be summarized as follows.
\begin{itemize}
\item The fiber off-pointing strategy brings a contrast improvement of up to a factor of six for dual-field observations of companions below 80~mas separation. The implementation of this technique is straightforward.
\item The inner working angle has become narrower since the upgrade of the GRAVITY FT. Below 45~mas, fitting the stellar speckles with a third-order polynomial gives better results than the fourth or sixth-order polynomials previously recommended.
\item Our detection of the brown dwarf Gaia~...6464~B shows how the combination of Gaia and GRAVITY leads to a precise measurement of the dynamical masses of the companion  ($78.34_{-2.50}^{+2.62}$\,\SI{}{M_{\mathrm{Jup}}}) and the primary ($0.53_{-0.02}^{+0.02}$\,\SI{}{M_\odot}). The detection of the brown dwarf is an archetypal example of a GRAVITY observation at the edge of the inner working angle.
\item In dual-field mode, GRAVITY can observe companions at contrasts  of $4\times 10^{-5}$ ($\Delta \mathrm{K}= 11$~mag) down to a separation of 75~mas. Due to the limited sampling of the UV plane on the UT, the detection limits below 100~mas are strongly affected by the relative orientation between the primary/secondary and the longest baselines. 
\item Observing the same target for about 3h pushes the detection limit down to $2\times 10^{-5}$ ($\Delta \mathrm{K}= 11.7$~mag) at 100~mas. The limiting factor appears to be speckle structures whose amplitude scales with the coherent flux and that slowly average over time.
\end{itemize}
These findings are the result of experience gained over 5 years of exoplanet and brown dwarf observations with GRAVITY. A better understanding of the instrument enabled us to identify promising avenues for improving observations and to consider possible synergies with other instruments.

\subsection{Comparison with other direct imaging instruments}
Figure~\ref{fig:DI_curves} compares the contrast curve of GRAVITY estimated in this paper with those of some other exoplanet-imaging instruments. This paper focuses on companion observations with GRAVITY below 300 mas separation. However, it should be noted that the off-axis mode of the instrument allows observations at separations of up to 2 arcsec on the UT. For separations of greater than 200~mas, the detection capabilities of GRAVITY are comparable to single-telescope imagers. Indeed, all exoplanets directly imaged with SPHERE at the VLT have also been detected with GRAVITY. GRAVITY is especially useful because of its 50~$\mu$as precision in relative astrometry and its higher spectral resolution. For long-period orbits, the accurate astrometry of GRAVITY can constrain orbital parameters within a few years. Moreover, these accurate astrometric observations will adequately complement further observations at the same level of accuracy as should be achieved by the E-ELT, providing a large time baseline.

Closer than 200~mas, GRAVITY is unique because this region is mostly beyond the reach of current single-telescope instruments. For instance, $\beta$\ Pictoris\ c and HD 206893 c are easily detected by GRAVITY but have so far remained undetected by SPHERE or GPI. We note that the molecular mapping technique is being increasingly used for exoplanet detection and characterization \citep{2023arXiv230912390V}. Molecular mapping has formally no self-subtraction inner working angle (unlike the spectral or angular differential deconvolutions). Instead, the method relies on the signal provided by sharp spectral features in the companion spectra, and is therefore highly photon consuming and requires longer integration times; it is blind to the continuum part of the spectrum, including the absolute flux of the companion. This is complementary to GRAVITY observations at moderate spectral resolution. The question of whether or not the molecular mapping technique can be applied to GRAVITY data themselves remains unanswered.

\subsection{Synergy with Gaia}
Brown dwarf companion candidates are already available in Gaia DR3. The next Gaia DR4 will provide the individual epochs of the stellar proper motion and a catalog of possibly thousands of planetary-mass companions suitable for characterization from the ground.

\begin{figure}[t]
    \centering
    \includegraphics[width=0.98\columnwidth]{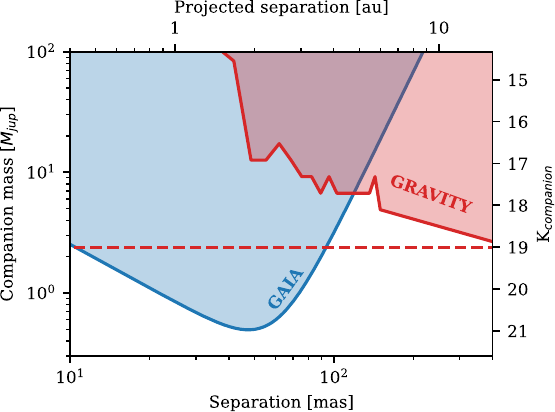}
    \caption{Comparison of the sensitivity of Gaia and GRAVITY for a K=6~mag star at 40 pc. The Gaia sensitivity curve is adapted from \cite{sozzetti2010}. The scaling of the K-magnitude with the mass is estimated from the observations of Gaia~...6464~B analyzed in this paper and previous companion observations from the ExoGRAVITY program. The dashed red line shows the current K-magnitude limit in the SC of GRAVITY.}\label{fig:gaiaGravity}
\end{figure}

Figure~\ref{fig:gaiaGravity} combines the GRAVITY detection limits derived in this paper with the expected Gaia sensitivity to substellar companions \citep{sozzetti2010}. This comparison shows that the GRAVITY sensitivity already overlaps with the Gaia sensitivity to companions in the 50 to 100 mas separation range around the nearest stars. Thus, GRAVITY bridges the gap for direct detection of these Gaia candidates before the E-ELT instruments \citep{houlle2021}. We note that Figure~\ref{fig:gaiaGravity} uses the favorable case of a star located at 40~pc to convert the linear and angular separation. For more distant stars, the Gaia sensitivity peak shifts toward even shorter angular separation. This highlights the interest of the unique inner working angle of GRAVITY and emphasizes the importance of pushing the instrument to even shorter separations and deeper contrast.

\subsection{Perspectives with GRAVITY+}
The GRAVITY+ upgrade includes a faint mode that turns off the metrology lasers during the acquisitions to provide better S/N for faint targets \citep{widmann2022} and a new extreme AO at the Coudé focus of the four UTs \citep{eisen2019}. The replacement of the 20~year-old MACAO will significantly improve the wavefront correction for atmospheric turbulence and enable correction of instrument internal aberrations (not possible with the current AO). The expected performance is a Strehl ratio of about 0.8 on natural guide stars, compared with the ratio of $\sim$0.3 presently provided by MACAO. This fact alone will already contribute to limiting the starlight injection at the planet's position by a factor 3 to 4. However, it is possible to go even further. The higher level of AO performance will allow wavefront control techniques to further reduce the starlight at the fiber position when observing the companion \citep{pourre2022}. The technique consists in injecting offset modes in the AO to tackle diffraction and static aberrations that couple into the SC fiber. Assuming a telescope close to the diffraction limit, and together with the fiber position offset presented in this paper, this technique could reduce the speckle amplitude by up to two orders of magnitude at separations of from 60 to 140~mas. In real conditions on-sky, \cite{xin2023}  showed that wavefront control on KPIC can result in up to a factor 3 reduction in K-band stellar flux injected into a single-mode fiber at 2~$\lambda/D$. It indicates that, overall, the developments at VLTI and GRAVITY might push the contrast limits to $3\times10^{-6}$ ($\Delta \mathrm{K}= 13.8$~mag) down to 60~mas separation. For the closest stars in our galactic neighborhood, this will enable observations of the young Jupiter-mass planets, possibly down to the snowline at 2 to 5~au from their star. % This will contribute to a better understanding of the formation and evolution of planetary systems.

\begin{acknowledgements}
This work is based on observations collected at the European Southern Observatory under ESO programmes 1104.C-0651, 0110.C-0182, 60.A-9102 and 0112.C-2396(C). We would like to thank the Paranal staff, especially the engineers, technicians and astronomers on the VLTI.
    This work has also made use of data from the European Space Agency (ESA) mission
{\it Gaia} (\url{https://www.cosmos.esa.int/gaia}), processed by the {\it Gaia}
Data Processing and Analysis Consortium (DPAC,
\url{https://www.cosmos.esa.int/web/gaia/dpac/consortium}). Funding for the DPAC
has been provided by national institutions, in particular the institutions
participating in the {\it Gaia} Multilateral Agreement.
The authors acknowledge the support of the French Agence Nationale de la Recherche (ANR), under grant ANR-21-CE31-0017 (project ExoVLTI). DD acknowledges the support from the ERC under the European Union’s Horizon 2020 research and innovation program (grant agreement CoG – 866070). GDM acknowledges the support of the DFG priority program SPP 1992 ``Exploring the Diversity of Extrasolar Planets'' (MA~9185/1) and from the Swiss National Science Foundation under grant 200021\_204847 ``PlanetsInTime''. Parts of this work have been carried out within the framework of the NCCR PlanetS supported by the Swiss National Science Foundation. J.J.W., A.C., and S.B. acknowledge the support of NASA XRP award 80NSSC23K0280. This research has made use of the NASA Exoplanet Archive, which is operated by the California Institute of Technology, under contract with the National Aeronautics and Space Administration under the Exoplanet Exploration Program. This research has also made use of the Jean-Marie Mariotti Center \texttt{Aspro}
service \footnote{Available at \url{http://www.jmmc.fr/aspro}}.
This research has also made use of the following python packages: matplotlib \citep{hunter2007}, numpy \citep{harris2020}, hcipy \citep{por2018hcipy} and astropy \citep{astropy2018}.
\end{acknowledgements}


\bibliographystyle{aa} % style aa.bst
\bibliography{bibli.bib} % your references Yourfile.bib

\begin{appendix}
\onecolumn
\section{Complex visibilities for a planet detection on GRAVITY} \label{sec:ExoSIGNAL}
\begin{figure}[ht]
\centering
\begin{subfigure}{0.9\textwidth}
  \centering
  \includegraphics[width=0.7\linewidth]{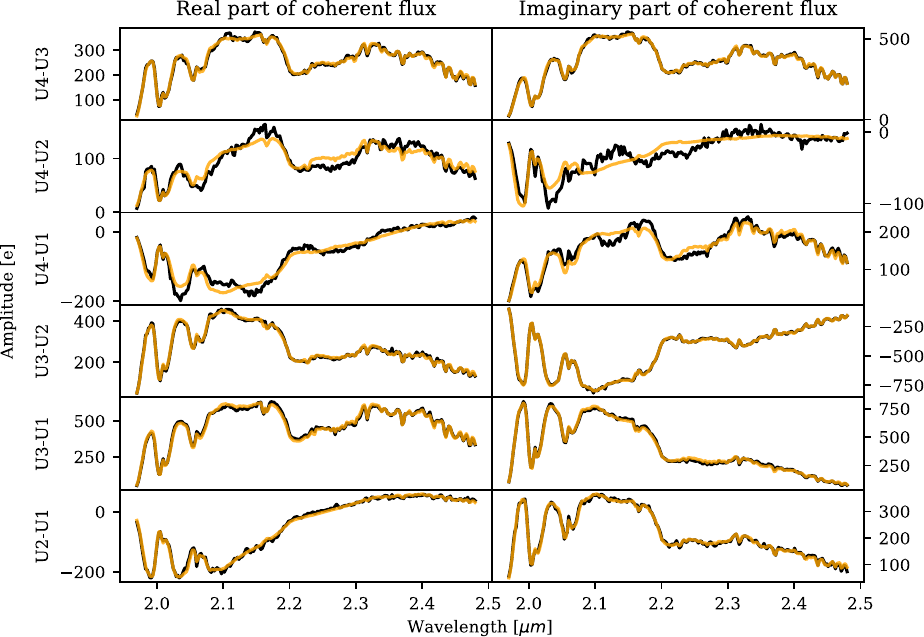}
  \caption{Speckles fit}
  \label{fig:stafit}
\end{subfigure}%
\newline
\begin{subfigure}{0.9\textwidth}
  \centering
  \includegraphics[width=0.7\linewidth]{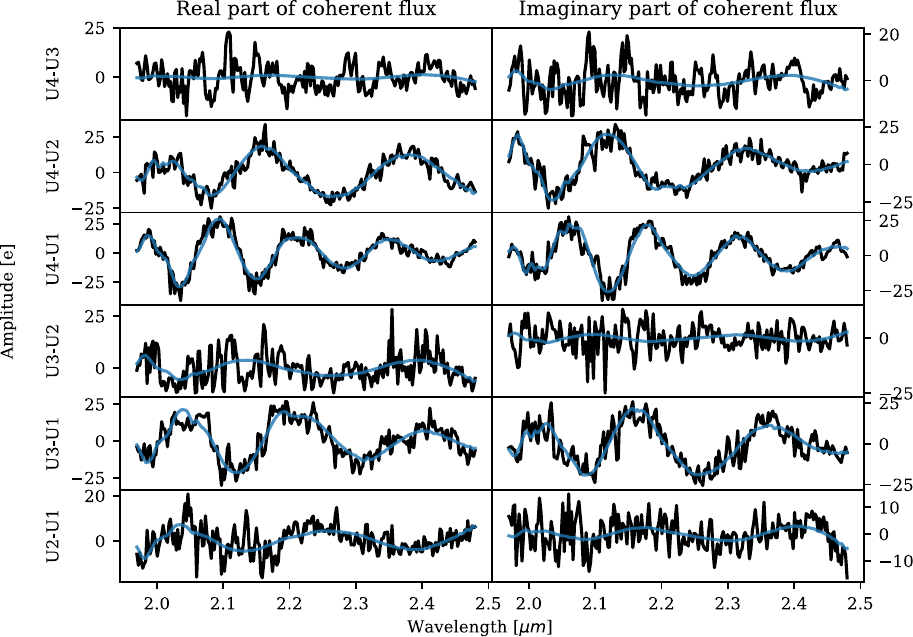}
  \caption{Planet fit}
  \label{fig:plafit}
\end{subfigure}
\caption{Example of speckle fit (a) and planet fit (b) in GRAVITY complex visibilities for an injected companion at 68~mas separation and $8\times 10^{-4}$ contrast. (a) Black: Complex visibilities $V_{\textrm{oncompanion}}$ as outputted by the general GRAVITY pipeline and phase referenced on the star. Orange: Speckles fit with the polynomial modulation corresponding to the first term in Eq.~(\ref{eq:exograv2}). (b) Black: $V_{\textrm{oncompanion}}$ with the speckle fit subtracted. Blue: Planet fit, corresponding to the second term in Eq.~(\ref{eq:exograv2}) }
\label{fig:ExoSIGNAL}
\end{figure}

Figure~\ref{fig:ExoSIGNAL} shows how we retrieve the companion signal buried in the speckle flux. 
The ExoGRAVITY reduction script performs a joint fit of the speckles and the planet on the complex visibilities. The correlated and uncorrelated noises can be seen in Fig.~\ref{fig:plafit}, especially on the U4-U3 and U3-U2 baselines, that provide a weak planet signal.

\clearpage
\twocolumn
\section{ExoGRAVITY observations used for detection limits} \label{sec:ExoObs}
For the companion injection and retrieval, we used archival ExoGRAVITY observations. We chose the observations based on the criteria of the relative separation of the SC fiber from the star, and good atmospheric conditions. Data sets are detailed in Table~\ref{tab:exoobs}.

\begin{table}[ht]
\centering
\caption{Details of the ExoGRAVITY observations used for the companion injection and retrievals.}

\begin{tabular}{c c c c c} 
 \hline
   Injection separation limits & 30 to 45~mas & 45 to 70~mas \\  
 \hline
  SC fiber separation [mas] & 54 & 72 \\
 Target star & HD 17155 & HD 206893\tablefootmark{a} \\
 Date & 2022-08-19 & 2021-08-28  \\
 K star [mag] & 6.5 & 5.6  \\
 Seeing [arcsec] & 0.54 & 0.62 \\ 
 $\tau_0$ [ms] & 9-18 & 2-8  \\
 Integration time & $5\times 32 \times 10$~s & $5\times 32 \times 10$~s \\
  \hline
   Injection separation limits & 70 to 100~mas & 100 to 130~mas   \\  
 \hline
   SC fiber separation [mas] & 92 & 111 \\
 Target star & $\beta$ Pictoris & HD 206893 \\
 Date & 2021-01-06 & 2021-10-17 \\
 K star [mag] & 3.5 & 5.6 \\
 Seeing [arcsec]  & 0.45 & 0.50 \\ 
 $\tau_0$ [ms]  & 8-12 & 2-4 \\
  Integration time & $5\times 32 \times 10$~s & $5\times 32 \times 10$~s \\
    \hline
   Injection separation limits & 130 to 150 mas & 320~mas   \\  
 \hline
   SC fiber separation [mas] & 136 &  320 \\
 Target star & $\beta$ Pictoris & AF Leporis \\
 Date & 2020-02-10 & 2023-12-23 \\
 K star [mag] & 3.5 & 4.9 \\
 Seeing [arcsec]  & 0.81 & 0.52 \\ 
 $\tau_0$ [ms]  & 6-15 & 5-10\\
  Integration time & $5\times 32 \times 10$~s & $5\times 12 \times 30$~s \\
\end{tabular}
\tablefoot{\tablefoottext{a}{The observations of HD 206893 on 2021-08-28 were performed in a search for the HD~206893~c planet and no planet was detected in this data set.}}
\label{tab:exoobs}
\end{table}

\clearpage

\end{appendix}

\end{document}